\documentclass[reprint,superscriptaddress,amsmath,amssymb,
aps,prb]{revtex4-2}
\usepackage{hyperref}
\hypersetup{colorlinks,allcolors=blue}
\usepackage{dcolumn}
\usepackage{bm}
\usepackage{tabularx}
\newcolumntype{Y}{>{\centering\arraybackslash}X}
\usepackage{wrapfig}
\usepackage{graphicx} 
 \usepackage{bm}
 \usepackage{physics}
 \usepackage{multirow}
 \usepackage[dvipsnames]{xcolor}

\graphicspath{{pictures/}}

\begin{document}

\title{Josephson-like magnetic tunnel junction - transition from classical to quantum regime}
\author{V.V.~Yurlov}
\email{yurlov.vv@phystech.edu}
\affiliation{Moscow Center for Advanced Studies, Kulakova str. 20, Moscow, 123592, Russia}  
\affiliation{New Spintronic Technologies, Bolshoy Bulvar 30, bld. 1, 121205  Moscow, Russia}
\author{P.N.~Skirdkov}
\affiliation{New Spintronic Technologies, Bolshoy Bulvar 30, bld. 1, 121205  Moscow, Russia}
\affiliation{Prokhorov General Physics Institute of the Russian Academy of Sciences, Vavilova 38, 119991 Moscow, Russia}
\author{K.A.~Zvezdin}
\email{k.zvezdin@nst.tech}
\affiliation{New Spintronic Technologies, Bolshoy Bulvar 30, bld. 1, 121205  Moscow, Russia}
\author{A.K.~Zvezdin}
\affiliation{New Spintronic Technologies, Bolshoy Bulvar 30, bld. 1, 121205  Moscow, Russia}

\bibliographystyle{apsrev4-2}
\date{\today}

\begin{abstract}

We theoretically propose and analyze a Josephson-like magnetic tunnel junction (MTJ) structure that exhibits quantum spin dynamics analogous to those in superconducting Josephson junctions. By exploiting the isomorphism between the equations of motion for low-dissipation MTJs with easy-plane anisotropy and the Josephson phase dynamics, we construct a theoretical framework for realizing spintronic qubits. Within this framework, we identify the physical parameters—such as anisotropy constants, Gilbert damping, spin current amplitude, and geometric factors—that govern the transition from classical to quantum behavior. 
We show that different types of spintronic qubits, including analogs of charge, flux, and transmon superconducting qubits, can be implemented depending on the hierarchy of energy scales. A Hamiltonian formalism is developed for each regime, enabling an analytical treatment of the two-level quantum dynamics and estimation of coherence times. In particular, we demonstrate that the spin current can be used not only to excite but also to stabilize the qubit states through dissipation control. These findings provide a route toward integrating spintronic qubits into CMOS-compatible architectures and lay the groundwork for a fully spintronic platform for quantum computation.

\end{abstract}

\maketitle
\section{Introduction}
The pursuit of quantum computing solutions capable of addressing problems beyond the capabilities of classical systems has become increasingly urgent~\cite{https://doi.org/10.1002/1521-3978(200009)48:9/11<771::AID-PROP771>3.0.CO;2-E, doi:10.1098/rsta.2003.1227, doi:10.1126/science.273.5278.1073}. Superconducting circuits based on Josephson junctions currently form the backbone of quantum logic in many platforms~\cite{doi:10.1063/1.2780165,10.1063/1.5089550,10.1063/1.2155757}, where nonlinearity introduced by the junction allows for the formation of discrete energy levels—enabling two-level quantum systems, or qubits.
\par
Despite their success, Josephson-based qubits are constrained by their requirement for ultra-low temperatures, typically in the millikelvin range. This limitation has driven the search for alternative platforms. One particularly promising candidate is spintronics, where magnetic structures exhibit quantum dynamics analogous to those of Josephson junctions~\cite{Zvezdin2002}.
\par
Quantum Josephson-like regimes have been theoretically and experimentally demonstrated in a variety of spin-based systems, including skyrmions~\cite{PhysRevLett.127.067201}, chiral domain walls~\cite{PhysRevResearch.5.033166}, and Bose-Einstein magnon condensates~\cite{Demokritov2006,PhysRevLett.102.187205,Bozhko2016,Divinskiy2021,Rüegg2003,vetoshko2020bose,bunkov2020features,petrov2024transition}, whose mathematical frameworks often mirror superconducting qubit models. In particular, antiferromagnetic films—with their high resonance frequencies—are attractive candidates for spin-based quantum logic~\cite{Khymyn2017,PhysRevB.85.014523,PhysRevB.94.094434,PhysRevB.104.104402}. Some of these systems even demonstrate Josephson-like effects at room temperature~\cite{Khymyn2017, Zvezdin2002}.
\par
Magnetic tunnel junctions (MTJs), composed of two ferromagnetic layers separated by an insulating barrier, offer a particularly promising implementation. These structures exhibit high magnetoresistance and are already widely used in memory cells~\cite{PhysRevApplied.13.034035,Khvalkovskiy_2013}, nano-oscillators~\cite{Kiselev2003,PhysRevLett.92.027201,PhysRevApplied.22.024019}, spectrum analyzers~\cite{10.1063/1.5044435}, neuromorphic computing~\cite{Romera2018}, random-bit generators~\cite{Fukushima_2014,montoya2019magnetization,jenkins2019nanoscale}, and spintronic rectifiers~\cite{tulapurkar2005spin,skirdkov2020spin,finocchio2021perspectives}. The magnetic anisotropy of the free layer — either perpendicular (PMA)~\cite{doi:10.1063/1.3057974, Ikeda2010,RevModPhys.89.025008} or in-plane (EPA) — plays a critical role in defining these dynamics. In particular, EPA-based MTJs can exhibit spin dynamics that are isomorphic to those of Josephson junctions~\cite{Zvezdin2002}, with the magnetostatic anisotropy playing the role of the Josephson energy.
\par
This analogy between EPA-MTJ dynamics and Josephson junction physics paves the way for a new class of spintronic quantum devices. While the feasibility of Josephson-like spin dynamics in MTJs has been previously established~\cite{Zvezdin2002}, the physical conditions under which such systems behave as true quantum two-level systems remain to be precisely defined. Importantly, MTJs are CMOS-compatible and already industrially deployed, making them ideal candidates for scalable quantum integration within existing semiconductor infrastructure.
\par
In this work, we propose a Josephson-like quantum MTJ system driven by spin current and systematically investigate the conditions under which quantum spin dynamics can be achieved. We first develop an analytical model of classical dynamics based on the Landau-Lifshitz-Gilbert (LLG) equation and derive an effective Lagrangian under low-damping conditions, showing its correspondence with Josephson junction theory. We then identify the energy scales and system parameters that enable a quantum regime and construct a Hamiltonian framework to describe analogs of transmon, flux, and charge qubits within MTJ structures.
\par
Finally, we evaluate key qubit metrics — including coherence time, anharmonicity, and critical temperature — for different geometries and material parameters. We demonstrate how spin currents can be used to modulate the effective damping and thus control coherence, expanding the practical utility of MTJ-based quantum devices. These results support the potential of MTJs as a viable platform for scalable, solid-state spintronic qubits.

\section{Model and basic equation}

\begin{figure}[h!]
\begin{center}
\includegraphics[width=0.8\linewidth]{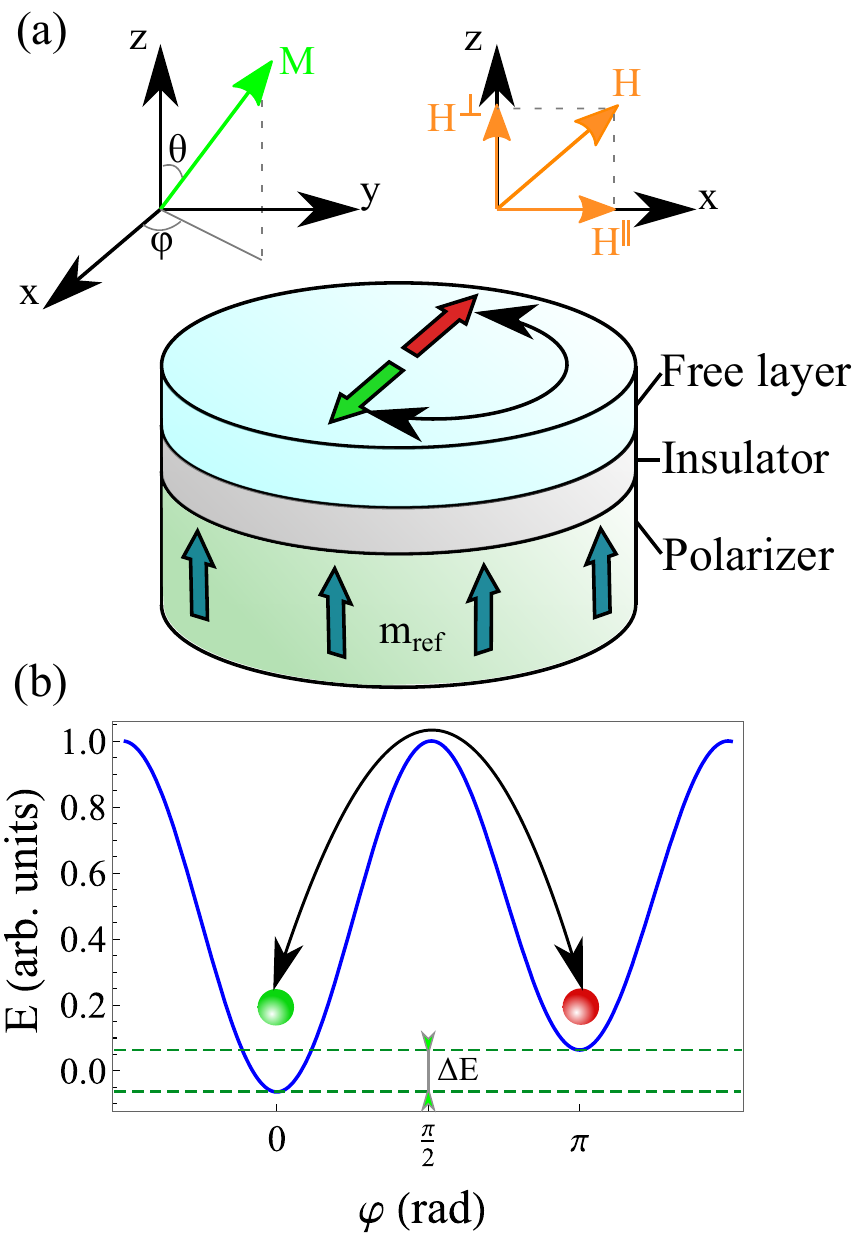}
\caption{
(a) Schematic representation of the MTJ structure. $H$ is the external magnetic field, $M$ is the magnetization of the free layer, and $\varphi$ and $\theta$ are the azimuthal and polar angles, respectively. 
(b) Potential energy of the magnetic layer as a function of the azimuthal angle at a fixed external magnetic field $H = 100$~Oe and $\theta \approx \pi/2$.
}
\label{Fig1}
\end{center}
\end{figure}

To begin with, it is necessary to provide a general description of the problem. We consider an easy-plane-based MTJ structure consisting of a ferromagnetic polarizer, a spacing insulator, and a free magnetic layer. This is illustrated in Fig.~\ref{Fig1}(a). We introduce a coordinate system with the $x$ and $y$ axes in the film's plane, and the $z$ axis perpendicular to the sample surface. The MTJ is exposed to an external magnetic field $\textbf{H}$, which lies in the $x$-$z$ plane (see Fig.~\ref{Fig1}(a)). A characteristic feature of this magnetic film is the presence of weak in-plane anisotropy, similar to the case of orthorhombic anisotropy. In this geometry, we obtain a two-well potential energy profile (see Fig.~\ref{Fig1}(b)), with the possibility of realizing a quantum two-level system under certain MTJ parameters. 
\par
In the first part of this paper, we discuss the classical magnetization dynamics induced by a short pulse of spin current, and construct a phenomenological Lagrangian formalism for the problem. In the next section, we use dynamic equations in the low-dissipation approximation to show how a quantum approach, similar to superconducting Josephson junction theory~\cite{doi:10.1063/1.2780165}, can be applied to this system. We determine the MTJ parameters and physical conditions required for implementing an MTJ-based Josephson-like qubit.
\par

Spin dynamics in the magnetic film is described by the Landau-Lifshitz-Gilbert (LLG) equation:

\begin{equation}\label{eq1}
\begin{gathered}
    \dfrac{d \textbf{m}}{dt} = -\dfrac{\gamma}{M_s}\textbf{m}\times \dfrac{\delta E}{\delta\textbf{m}}+ \alpha \textbf{m} \times \dfrac{\partial \textbf{m}}{\partial t} - j\cdot\textbf{m}\times\textbf{m}_{\mathrm{ref}}\times\textbf{m}, \\
    E = K_1\cos^2\theta + K_2\sin^2\theta\sin^2\varphi -  M_{\mathrm{s}}H^{\perp}\cos\theta - \\ - M_{\mathrm{s}}H^{\parallel}\sin\theta\cos\varphi,
\end{gathered} 
\end{equation}

where $M_\mathrm{s}$ is the saturation magnetization of the free magnetic layer, $\textbf{m}$ is the unit vector aligned with the magnetization vector, $\alpha$ is the Gilbert damping parameter, $\gamma$ is the gyromagnetic ratio, and $K_{1,2}$ are the magnetostatic anisotropy constants. Throughout this paper, we assume $K_1 \gg K_2 > 0$, corresponding to the demagnetization factors. $H^{\perp}$ and $H^{\parallel}$ are the $z$ and $x$ components of the external magnetic field, respectively, and $\theta$ and $\varphi$ are the polar and azimuthal angles.
The spin torque term includes 
$j = \dfrac{\gamma \hbar J P}{2 e M_\mathrm{s} d}$, 
where $J$ is the electric current density, $e > 0$ is the electron charge, $d$ is the thickness of the magnetic layer, $\textbf{m}_{\mathrm{ref}}$ is the unit magnetization vector of the ferromagnetic polarizer, and $P$ is the spin current polarization.
\par
Without loss of generality, we consider only the Slonczewski-like torque, since the field-like component merely renormalizes the effective magnetic field. It is worth noting that, in this work, the torque arises from spin-transfer effects associated with charge current injection. However, this mechanism is not optimal for qubit applications due to Joule heating. An alternative is the use of pure spin currents, which can be generated via the spin Hall effect, Rashba-Edelstein effect, or more generally at interfaces between a ferromagnet and materials with strong spin-orbit interaction (e.g., heavy metals or topological insulators). These mechanisms can induce torques of similar form~\cite{garello2013symmetry}, without requiring charge current flow through the ferromagnet. Thus, the torque form used in Eq.~(\ref{eq1}) retains generality in the context of spintronic qubit modeling.
\par

First, let us discuss the features of the classical magnetization dynamics for the considered structure. In the spherical coordinate system, Eq.~(\ref{eq1}) can be written as:
\begin{equation}\label{eq2}
\begin{cases}
\begin{gathered}
\dot{\theta}\sin\theta +\alpha\dot{\varphi}\sin^2\theta = -\dfrac{\omega_2}{2}\sin^2\theta\sin2\varphi - \\ - \omega^{\parallel}_H\sin\theta\sin\varphi + T_{\varphi}\\
\alpha\dot{\theta} - \dot{\varphi}\sin\theta = \dfrac{1}{2}(\omega_1 - \omega_2\sin^2\varphi)\sin2\theta - \\ - \omega^{\perp}_H\sin\theta + \omega^{\parallel}_H\cos\theta\cos\varphi + T_{\theta}
\end{gathered}  
\end{cases},
\end{equation}
where $\omega_{1,2} = \gamma \dfrac{2 K_{1,2}}{M_s}$, $\omega^{\parallel}_H = \gamma H^{\parallel}$, and $\omega^{\perp}_H = \gamma H^{\perp}$. The terms $T_{\theta}$ and $T_{\varphi}$ denote the components of the spin-transfer torque. These components depend on the orientation of the polarizer $\textbf{m}_{\mathrm{ref}}$ and, in the spherical coordinate system, can be expressed as:
\begin{itemize}
  \item for the $x$-direction: $T_{\theta} = -j \sin\varphi$, \quad $T_{\varphi} = -j \cos\varphi \sin\theta \cos\theta$,
  \item for the $y$-direction: $T_{\theta} = j \cos\varphi$, \quad $T_{\varphi} = -j \sin\varphi \sin\theta \cos\theta$,
  \item for the $z$-direction: $T_{\theta} = 0$, \quad $T_{\varphi} = j \sin^2\theta$.
\end{itemize}
To obtain equations of magnetization dynamics that are mathematically analogous to Josephson junction theory, we assume $\textbf{m}_{\mathrm{ref}} = (0, 0, 1)$ (see Fig.~\ref{Fig1}(a)). However, as will be discussed later, using an in-plane polarizer provides certain advantages for MTJ-based qubits.
The potential energy of the magnetic layer forms a double-well profile, with an energy gap $\Delta E$ between stable states, as shown in Fig.~\ref{Fig1}(b). By minimizing the potential energy with respect to the angles $\theta$ and $\varphi$, we find two stable states corresponding to the angle pairs $(\varphi^{(1)} = 0,\; \theta^{(1)} = \pi/2)$ and $(\varphi^{(2)} = \pi,\; \theta^{(2)} = \pi/2)$ at zero external field ($H = 0$).
\par

The spin current flowing through the magnetic film excites spin dynamics and drives the magnetization out of equilibrium. It is known that under weak dissipation, low temperatures, and small magnetic volume, quantum effects can emerge in magnetic structures. To describe this specific regime, we adopt a Lagrangian formalism under the low-dissipation approximation $\omega_d \ll \omega_r$, where $\omega_r = \sqrt{\omega_1 \omega_2}$ is the resonance frequency, and $\omega_d = \alpha \omega_1$ is the dissipation frequency.
\par

Assuming small damping and introducing the substitution $\theta = \pi/2 - \widetilde{\theta}$ with $\widetilde{\theta} \ll 1$, Eq.~(\ref{eq2}) can be rewritten as:
\begin{equation}\label{eq3}
\begin{cases}
\begin{gathered}
\dot{\widetilde{\theta}} = \dfrac{1}{2}\omega_2\sin2\varphi + \omega^{\parallel}_{H}\sin\varphi - j(t) \\
\dot{\varphi} = -\omega_1\widetilde{\theta} + \omega^{\perp}_{H} -\omega^{\parallel}_{H}\widetilde{\theta}\cos\varphi \\
\end{gathered}  
\end{cases}.
\end{equation}
Eliminating the angle $\widetilde{\theta}$ from (\ref{eq3}), we obtain a dynamic equation for the problem of spin dynamics, induced by a short current pulse: 
\begin{equation}\label{eq4}
\begin{cases}
\begin{gathered}
\ddot{\varphi} + \dfrac{\omega^2_r}{2}\sin2\varphi + \omega^{\parallel}_H\omega_1\sin\varphi = j(t)\omega_1\\
\varphi(0) = \varphi^{(1)}\\
\dot{\varphi}(0) = \omega_1\int\limits_0^\tau j(t)dt
\end{gathered}  
\end{cases}.
\end{equation}
\par
Equation~(\ref{eq4}) can be obtained under the assumption that $H^{\perp} \gg H^{\parallel}$, or equivalently, $\omega_1 \gg \omega^{\parallel}_H$. The initial angular velocity $\dot{\varphi}(0)$ can be derived from the dynamic equation~(\ref{eq4}) by integrating over the short duration of the spin current pulse $\tau$. The dissipation term can be included in Eq.~(\ref{eq4}) in the form $\alpha \omega_1 \dot{\varphi}$. Thus, Eq.~(\ref{eq4}) describes magnetization dynamics excited by an effective external force of the form $\sim j(t)\omega_1$ acting over a short time $\tau$.
\par
The Lagrangian and Hamiltonian corresponding to the magnetization dynamics described by Eq.~(\ref{eq4}) can be written as:
\begin{equation}\label{eq5}
\begin{gathered}
\mathcal{L}_{\mathrm{eff}} = \dfrac{m^*\dot{\varphi}^2}{2} - E_{\mathrm{eff}},\\
\mathcal{H}_{\mathrm{eff}} = \dfrac{P_{\varphi}^2}{2m^*} + E_{\mathrm{eff}}, 
\end{gathered}
\end{equation}
where $P_{\varphi} = \partial\mathcal{L}_{\mathrm{eff}}/\partial\dot{\varphi}$ is the generalized momentum, $m^* = M_s/(\gamma \omega_1)$ is the effective mass, and $E_{\mathrm{eff}}$ is the effective potential energy derived from Eq.~(\ref{eq4}). 
From the second equation in Eq.~(\ref{eq3}), it can be seen that the generalized momentum $P_{\varphi}$ can be expressed through the projection of the magnetization onto the $z$-axis, $M_z = M_s \cos\theta \approx M_s\widetilde{\theta}$, as:
\begin{equation}\label{eq21}
    P_{\varphi} = \frac{ M_s H^{\perp}}{\omega_1} - \frac{M_z}{\gamma}.
\end{equation}
As a result, the effective Lagrangian, Hamiltonian, and potential energy of the system described by Eq.~(\ref{eq5}) and Eq.~(\ref{eq21}) take the following form:
\begin{equation}\label{eq6}
\begin{gathered}
\mathcal{L}_{\mathrm{eff}} = \dfrac{M_s}{2\gamma\omega_1}\dot{\varphi}^2 - E_{\mathrm{eff}},\\
\mathcal{H}_{\mathrm{eff}} = \dfrac{\omega_1 M_s}{2\gamma}\Big(\dfrac{M_z}{M_s} - \gamma\dfrac{H^{\perp}}{\omega_1}\Big)^2 + E_{\mathrm{eff}},\\
E_{\mathrm{eff}} = - \dfrac{K_2}{2} \cos2\varphi - M_sH^{\parallel}\cos\varphi - \dfrac{j(t)M_s}{\gamma}\varphi.
\end{gathered}  
\end{equation}

\par
It can be seen that the mathematical form of the MTJ equations in Eq.~(\ref{eq6}), up to the substitution of variables $2\varphi \rightarrow \phi$, is isomorphic to the corresponding equations for superconducting Josephson junctions, which have the form $\mathcal{H} = E_c(n - n_g)^2 - E_J\cos\phi - \frac{\hbar}{2e}I_e\phi$, where $n$ is the number of Cooper pairs, $\phi$ is the phase difference across the Josephson junction, $E_c$ is the charging energy, $E_J$ is the Josephson energy, $I_e$ is the applied current, and $n_g$ is an externally controlled gate parameter~\cite{doi:10.1063/1.2780165}. Thus, we can identify the corresponding physical quantities in the two systems. 
\par
In 1963, Anderson~\cite{anderson1964lectures} proposed describing macroscopic quantum effects in the Josephson junction using macroscopic charge and phase as operators of generalized momentum and coordinate. In our case, the quantum coherent state is defined by the generalized momentum $P_{\varphi}$, which corresponds to the projection of the magnetic moment along the $z$-axis, and the azimuthal angle $\varphi$, respectively. Consequently, the quantum state of the MTJ-based qubit is described by a pair of azimuthal and polar angles on the Bloch sphere.
For the MTJ-based qubit, both anisotropy and Zeeman energies can play the role of Josephson energy due to their similar symmetry. The spin current corresponds to the bias current $I_e$, and the effective mass is analogous to the charging energy. Note that while the notation in Eq.~(\ref{eq4}) suggests that $\sqrt{\omega_1 \omega^{\parallel}}$ should act as the plasma frequency, we assume that $\omega_r$ plays the role of the zero-bias Josephson plasma frequency, since it does not depend on external parameters.
Therefore, by selecting appropriate parameters for the magnetic film, it becomes possible to analyze quantum effects in the MTJ using mathematical techniques developed for Josephson junction theory. Moreover, the Lagrangian and Hamiltonian in Eq.~(\ref{eq6}) can also be derived using the method of quantum coherent states~\cite{fradkin_2013} for ferromagnetic, ferrimagnetic, and antiferromagnetic layers. The resulting mathematical formulations in the low-temperature regime are similar across these material types, differing only by constant prefactors~\cite{Zvezdin2002}.
\par
By introducing operators for the main quantities in the following form $\hat{M}_z = -i\dfrac{g\mu_B N}{V} \dfrac{\partial}{\partial \varphi}$ and $\hat{\varphi} = \varphi$, which satisfy the commutation relation $[\hat{\varphi}, \hat{M}_z] = i g \mu_B N / V$, we can express the MTJ Hamiltonian from Eq.~(\ref{eq6}) in the following form:
\begin{equation}\label{eq7}
\begin{gathered}
\mathcal{\hat{H}} = 4E_1\Big(\hat{m}_z - m^{\perp}_H\Big)^2 - E_2\cos2\hat{\varphi} - \\  - E^{\parallel}_{H}\cos\hat{\varphi} - W(t)\hat{\varphi},
\end{gathered}
\end{equation}
where $\hat{m}_z = -i\partial/\partial\varphi$, $E_1 = \omega_1 M_s V / 8\gamma$, $m^{\perp}_H = \xi^{-1/2} \gamma H^{\perp} / \omega_1$, $E_2 = K_2 V / 2$, $E^{\parallel}_{H} = M_s H^{\parallel} V$, $W(t) = j(t) M_s V / \gamma$, and $M_s = g\mu_B N / V$, $\mu_B$ is the Bohr magneton, $g$ is the Landé $g$-factor, $V$ is the sample volume, and $N$ is the number of spins.
Equation~(\ref{eq7}) corresponds to the Hamiltonian of a superconducting qubit in Josephson junction theory~\cite{doi:10.1063/1.2780165}, which was written earlier. We observe that the Hamiltonian~(\ref{eq7}) can be decomposed into two terms $\hat{\mathcal{H}} = \hat{\mathcal{H}}_0 + \hat{\mathcal{H}}_T(t)$, where
\begin{equation}\label{eq22}
\begin{gathered}
\hat{\mathcal{H}}_0 = 4E_1(\hat{m}_z - m^{\perp}_H)^2 - E_2\cos2\hat{\varphi} - \\ - E^{\parallel}_{H}\cos\hat{\varphi}, \\
\hat{\mathcal{H}}_T(t) = - W(t) \hat{\varphi}.
\end{gathered}
\end{equation} 
The term $\hat{\mathcal{H}}_0$ is $2\pi$-periodic function, representing the stationary part of the Hamiltonian. In contrast, the term $\hat{\mathcal{H}}_T(t)$ acts as a time-dependent perturbation and can be interpreted as the analog of a displacement current applied to the system. The notation (\ref{eq22}) will be used later in the text to simplify the observation of stationary and non-stationary cases.

\section{Discussion}
\begin{figure}[h!]
\begin{center}
\center{\includegraphics[width=0.7\linewidth]{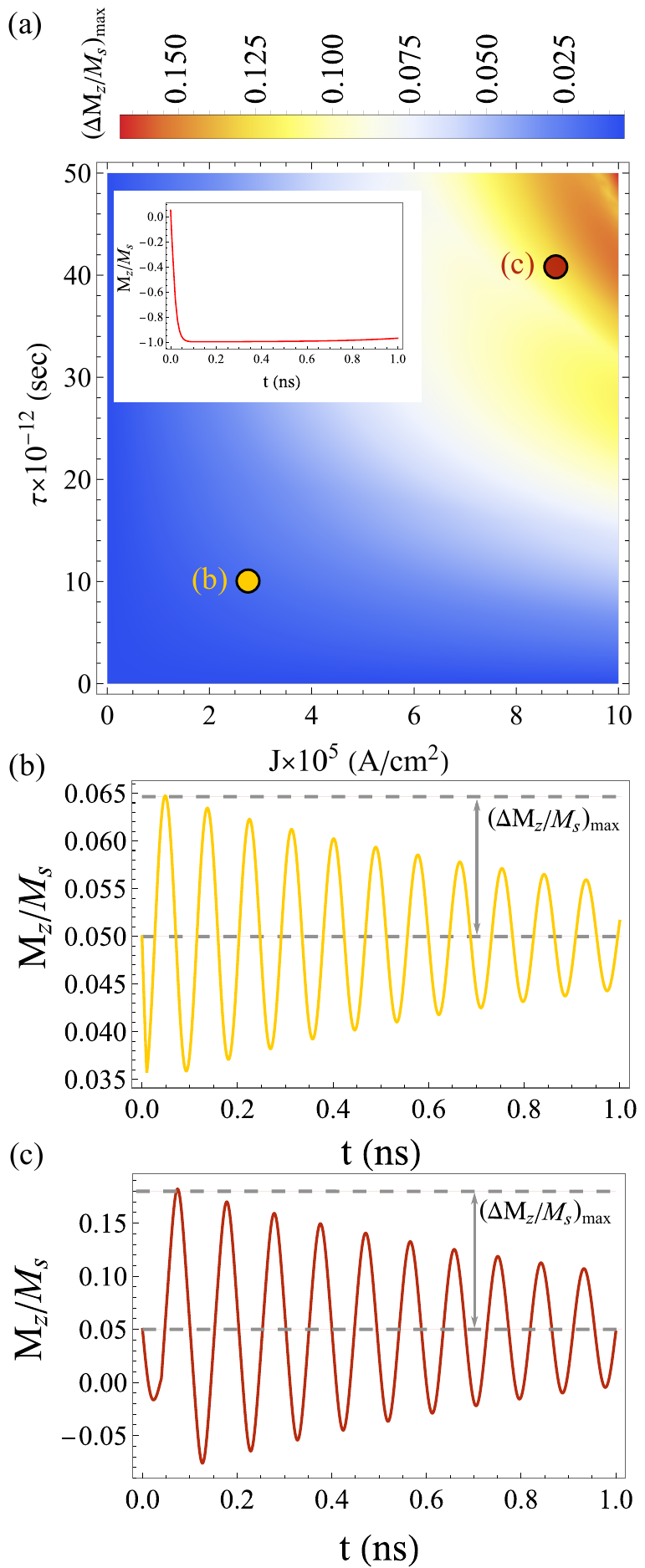}}
\caption{a) Diagram of the maximal value of the oscillations amplitude of the polar angle during the switching process in coordinates $\tau-J$. b) and c) are examples of the dynamic process for two different points of the diagram; red point corresponds to the $J = 9\cdot 10^5$ A/cm$^2$, $\tau = 40$ ps and green point corresponds to the $J = 3\cdot 10^5$ A/cm$^2$, $\tau = 10$ ps. Inset in (a) shows the saturated magnetization case  corresponds to the $J = 8\cdot 10^6$ A/cm$^2$, $\tau = 100$ ps.}
\label{Fig2}
\end{center}
\end{figure}
\begin{table*}
\caption{\label{Tab1}\textit{Possibility of the quantum approach}: characteristics of the magnetic tunnel junction at which the transition to the quantum regime is possible or impossible; "on" and "on/off" in the last column correspond to the fulfillment or partial fulfillment of the conditions for transition to the quantum regime. }
\small \centering
\renewcommand{\tabcolsep}{0.28cm}
\begin{ruledtabular}
\begin{tabular}{c|ccccccc}
\rule{0mm}{12pt}
$M_s$, G & $ D = d/2a $  & $d$, nm              & $\omega_1 = \gamma M_s N_z$, rad/s       & $\omega_2 = \gamma M_s N_y$, rad/s       & $\omega_r = \sqrt{\omega_1\omega_2}$, rad/s & $\alpha$   & Realisation \\ \hline
\rule{0mm}{12pt}
\multirow{4}{*}{500}         

                              &$ 0.01 $ & $ 0.2 $& \textcolor{Green}{$\approx 1.04 \cdot10^{11}$} & \textcolor{Green}{$\approx 4.60\cdot10^{9} $}  & $\approx 0.22 \cdot 10^{11}$                                & \textcolor{Green}{$\ll 0.21$} & \textcolor{Green}{On}          
                              \\
\rule{0mm}{12pt}               
                            &$ 0.05 $ & $ 1 $& \textcolor{Green}{$\approx 0.91 \cdot10^{11}$} & \textcolor{Green}{$\approx 0.15\cdot10^{11} $}  & $\approx 0.37 \cdot 10^{11}$                                & $\textcolor{Green}{\ll 0.41}$ & \textcolor{Green}{On}          
                              \\
\rule{0mm}{12pt}             
                              &$ 0.1 $ & $ 2 $& \textcolor{blue}{$\approx 0.79 \cdot10^{11}$} & \textcolor{blue}{$\approx 0.23\cdot10^{11} $}  & $\approx 0.43 \cdot 10^{11}$                                & \textcolor{Green}{$\ll 0.54$} & \textcolor{blue}{On/Off}          \\ 

\rule{0mm}{12pt}
                              & $0.2$ & 4 & \textcolor{blue}{$\approx 0.66 \cdot10^{11}$} & \textcolor{blue}{$\approx 0.33 \cdot10^{11}$} & $\approx 0.47 \cdot 10^{11}$                                &\textcolor{Green}{$ \ll 0.71$} & \textcolor{blue}{On/Off}       \\ 
\end{tabular}
\end{ruledtabular}
\renewcommand{\tabcolsep}{0.31cm}
\end{table*}
The theoretical results described above have certain limits of applicability, which should be discussed in more detail. We begin with the feasibility of using spin current as a trigger for quantum magnetization dynamics. A key assumption underlying the isomorphism with superconducting systems is that the polar angle remains close to $\theta \approx \pi/2$. Therefore, it is necessary to select suitable values for the pulse duration and electric current density to ensure this condition is maintained.
To this end, we numerically solve the equations~(\ref{eq2}) governing the classical dynamics of the magnetic moment, assuming low dissipation $\omega_d \ll \omega_r$ and $K_1 \gg K_2 > 0$. The following material parameters are used for the simulation: $P = 0.3$, $\omega_1 \approx 1 \cdot 10^{11}$~rad/s, $\omega_2 = 0.01\omega_1$, $K_1 \approx 1 \cdot 10^6$~erg/cm$^3$, $M_s = 500$~G, $H^{\parallel}/H^{\perp} = 0.1$, $H = 1000$~Oe, $d \sim 0.5$~nm, and $\alpha \sim 0.001$. A more detailed derivation of each parameter is provided in the next section (see Table~\ref{Tab1}).
The polarization value $P = 0.3$ is typical for current-generation tunnel junctions. However, recent progress—particularly with symmetry-filtered MgO-based MTJs—has enabled significantly higher polarizations, approaching the theoretical limit of $P = 1$, especially at low temperatures. Below, we retain $P = 0.3$ for analysis, with the understanding that advanced materials would require approximately three times lower current density. Additionally, lower Joule heating would help facilitate cryogenic operation.
\par
It is important to note that variations in the absolute value of the magnetic field $H$ do not significantly affect the diagram’s structure, but do change the ground state $\theta$, there by altering the deviation of the magnetization from the film plane. Therefore, to satisfy conditions $H^{\perp} \gg H^{\parallel}$ or $\omega_1 \gg \omega^{\parallel}_H$, and maintain $\theta \approx \pi/2$, the magnitude of $H^{\perp}$ should be limited to $\sim1000$~Oe, while $(H^{\parallel})_{\mathrm{max}} \sim 100$~Oe. For instance, under the given parameters, the equilibrium state yields $\cos\theta \approx 0.05$, which represents an acceptable deviation from the surface plane.
\par
To determine the regime in which quantum dynamics may be initiated, we construct a magnetization dynamics diagram in the $\tau$–$J$ coordinates, shown in Fig.~\ref{Fig2}(a). Here, we assume a square-shaped current pulse profile. The color scale indicates the maximum deviation of the polar angle from equilibrium, defined in the figure as $(\Delta M_z/M_s)_{\mathrm{max}}$. Using this diagram, one can select appropriate pulse durations and current densities that keep $\theta$ within the quantum-compatible regime. The acceptable operating range corresponds to the blue region, while the red region is excluded. In the red zone, the spin current oversaturates the magnetic structure, significantly displacing the magnetization vector from the film plane. This is illustrated in the inset of Fig.~\ref{Fig2}(a) for $J = 8\cdot 10^6$~A/cm$^2$ and $\tau = 100$~ps.
Figures~\ref{Fig2}(b) and \ref{Fig2}(c) present time-dependent profiles of the polar angle $\theta(t)$ for two representative points in the diagram. The green point corresponds to $J = 3\cdot 10^5$~A/cm$^2$, $\tau = 10$~ps, while the red point corresponds to $J = 9\cdot 10^5$~A/cm$^2$, $\tau = 40$~ps. The maximum deviation of the polar angle for these two points differs in ten times.
\par

\begin{table*}
\caption{\label{Tab2}Comparison table of the main parameters of Josephson superconducting transitions and MTJ structure. }
\small \centering
\renewcommand{\tabcolsep}{0.28cm}
\begin{ruledtabular}
\begin{tabular}{c|ccc}
\rule{0mm}{12pt}
\multirow{2}{*}{MTJ structure}
& Resonant frequency $\omega_r$, rad/s
                           &  Gilbert damping $\alpha$  & Temperature limitation $T$, K  \\ 

\rule{0mm}{12pt}
                               & $\sim 2\cdot10^{10}$
                           & $\sim 10^{-3}  - 10^{-5}$\cite{PhysRevLett.107.066604,10.1038/s41467-018-05732-1}  & $\sim 0.01- 1$  \\ \hline
                           \rule{0mm}{12pt}
\multirow{2}{*}{Josephson junction}
& Plasma frequency $\omega_{J}$, rad/s
                           &  Dissipation constant $\alpha_{J}$  & Temperature limitation $T$, K  \\ 

\rule{0mm}{12pt}
                               & $\approx 6\cdot10^{10}$ \cite{BLACKBURN20102827,BLACKBURN20161}
                           & $\sim 10^{-3} - 10^{-5}$  \cite{PhysRevLett.97.050502,BLACKBURN20102827,BLACKBURN20161}  & $\sim 0.01 - 0.3$  \cite{BLACKBURN20161,PhysRevLett.93.107002} \\
\end{tabular}
\end{ruledtabular}
\renewcommand{\tabcolsep}{0.31cm}

\end{table*}
\begin{table*}
\caption{\label{Tab3}\textit{Qubit type}: summarizing of the MTJ parameters for which Josephson-like qubit can be realized. Here we suppose that $N = 100$ and $H^{\parallel} = 100 $ Oe, $\beta  = 0.4$, $a = 10^{-6}$ cm, $M_s = g\mu_BN/V$, $E^{\parallel}_H \approx 1.85 \cdot 10^{-16}$ erg.}
\small \centering
\renewcommand{\tabcolsep}{0.28cm}
\begin{ruledtabular}
\begin{tabular}{cccccccc}
\rule{0mm}{12pt}
$V = d\cdot S$, cm$^3$ & $d$, nm & $E_1 = V \omega_1 M_s/8\gamma$, erg & $E_2 = V \omega_2 M_s/4\gamma$, erg  & $E_1/E_2$& $E_1/E^{\parallel}_H$ & $E_2/E^{\parallel}_H$ & Qubit type \\ \hline
\rule{0mm}{12pt}
                          $6.28\cdot10^{-21}$  & 0.1 &  $ \approx 8.33 \cdot 10^{-16} $& $\approx 4.09\cdot10^{-17}$  & $\approx 20.37$ & $\approx 4.49$ & $\approx 0.22$ & \textcolor{blue}{Charge}\\
\rule{0mm}{12pt}
                          $6.28\cdot10^{-20}$  & 1 &  $ \approx 7.06 \cdot 10^{-17} $& $\approx 2.36\cdot10^{-17}$  & $\approx 2.99$ & $\approx 0.38$ & $\approx 0.13$ & \textcolor{red}{Flux}\\

\rule{0mm}{12pt}
                          $2.51\cdot10^{-19}$  & 4 & $ \approx 1.29 \cdot 10^{-17} $& $\approx 1.28\cdot10^{-17}$  & $\approx 1.01$ & $\approx 0.07$& $\approx 0.06$ & \textcolor{red}{Transmon}\\
\end{tabular}
\end{ruledtabular}
\renewcommand{\tabcolsep}{0.31cm}
\end{table*}

The next step involves assessing the applicability of the operator formalism for macroscopic parameters. Given the mathematical structure of equation~(\ref{eq7}), we can describe the isomorphism between the considered magnetic structure and a superconducting Josephson junction. Based on this analogy, the following inequality must be satisfied in order to transition to a quantum description:
\begin{equation}\label{eq23}
kT \leq \hbar\omega_{01} \leq \Delta,
\end{equation}
where $T$ is the temperature of the magnetic layer, $\hbar\omega_{01}$ is the energy separation between the two lowest quantum states of the system, and $\Delta$ is the energy gap in the spin-wave spectrum. The left part of the inequality ensures suppression of thermal fluctuations, while the right part allows the variables $\varphi$ and $m_z$ to be treated as quantum operators.
In the simplest case, it can be shown that $\Delta \sim \hbar \omega_r$~\cite{Landau1978}, where $\omega_r$ is the resonance frequency. Therefore, the energy separation between the lowest levels should remain smaller than the magnon energy gap. However, this gap can be significantly increased by including higher-order anisotropy terms or by applying external magnetic fields.
Thus, when all the following conditions are satisfied—namely $\omega_d \ll \omega_r$, $K_1 \gg K_2 > 0$, $kT \leq \hbar\omega_{01} \leq \Delta$, and $\theta \approx \pi/2$—a quantum-mechanical treatment of the system becomes possible.
\par
The second step involves determining appropriate variable parameters for the MTJ. To this end, we rewrite the anisotropy constants as $K_1 = N_z M_s^2/2$ and $K_2 = N_y M_s^2/2$, where $N_z$ and $N_y$ are the demagnetizing factors. To complete the description of the model, we also introduce the third demagnetizing factor and the corresponding anisotropy energy $K_3 = N_x M_s^2/2$, with the constraint $N_x + N_y + N_z = 4\pi$. This form of the anisotropy energy allows us to consolidate all numerical data into a single Table~\ref{Tab1} by varying only the ratio of the demagnetizing factors. It should be noted, however, that the values of $K_1$ and $K_2$ may be altered by in-plane easy axes, additional magnetostatic contributions, or orthorhombic anisotropy.
For simplicity, we present all quantities assuming a fixed value of $M_s = 500$~G, to illustrate representative MTJ parameter values at different thicknesses. In the remainder of the text, we will vary the number of spins $N$ and the volume of the free layer $V$ instead of $M_s$. Note that in Table~\ref{Tab1}, the number of spins $N$ ranges from approximately $ \sim 100$ to $\sim 1000$.
\par
It is well known that demagnetization factors are determined by the geometric parameters of the structure. Therefore, the exact expressions for the demagnetizing factors are taken from Ref.~\cite{Beleggia_2005}. We assume that the structure is shaped as an elliptical cylinder, with the ratio of the major ($a$, along the x-axis) and minor ($b$, along the y-axis) semi-axes set to $\beta = b/a \approx 0.4$. For simplicity, we fix the major semi-axis at $a = 10^{-6}$~cm. By varying the thickness $d$ of the free magnetic layer, we can adjust the relative values of the demagnetizing factors. For the quantum regime considered here, the condition $N_z \gg N_y \gg N_x$ must be satisfied.
\par
Table~\ref{Tab1} illustrates the feasibility of achieving quantum states based on the relationship between $\omega_1 = \gamma M_s N_z$ and $\omega_2 = \gamma M_s N_y$, as well as the value of the dissipation constant $\alpha$. The table also accounts for the small value of $N_x$, which corresponds to a characteristic length $D = d / 2a$~\cite{Beleggia_2005} that ranges from $0$ to $0.2$. We select four representative values of $D$ in Table~\ref{Tab1} to capture general trends relevant to realization of the quantum state.
We observe that as $\omega_1$ decreases, the value of $\omega_2$ increases. However, the overall resonance frequency $\omega_r$ increases, indicating that the upper bound for the dissipation constant remains approximately $\alpha \sim 10^{-3}$ across all cases. Achieving such low $\alpha$ is challenging in practice, particularly for very thin magnetic layers. Nevertheless, it is known that certain ferromagnetic layers can exhibit damping parameters as low as $\approx 5 \cdot 10^{-3}$—for example, a 5~nm Py layer under low-temperature conditions~\cite{zhao2016experimental}. Note, that in the next paragraphs we will fix the number of spins $N = 100$. This assumption will lead to the decreasing in the absolute value of the $M_s$ and will tighten restrictions on dissipation. However, in that case the required values of the dissipation are still stay in the achievable range.  
\par
In our specific case, the restriction on the dissipation constant becomes less stringent with increasing magnetic layer thickness, making layers of 1--10~nm more favorable for experimental implementation. It is also worth noting that ultra-low damping values have been demonstrated in certain iron garnet heterostructures~\cite{PhysRevLett.107.066604,10.1038/s41467-018-05732-1}.
From Table~\ref{Tab1}, it is evident that for layer thicknesses $d = 0.2$~nm and $d = 1$~nm, all quantum criteria are satisfied, and these cases are marked as "On." In contrast, other thicknesses are labeled "On/Off" to reflect the fact that the feasibility of qubit realization is uncertain. This uncertainty follows from $\omega_1$ and $\omega_2$ becoming too close in value, despite more relaxed requirements for the damping constant. Nonetheless, we consider the full range of characteristic lengths in subsequent analysis. 
     
\par
\begin{figure}[h!]
\begin{center}
\center{\includegraphics[width=0.95\linewidth]{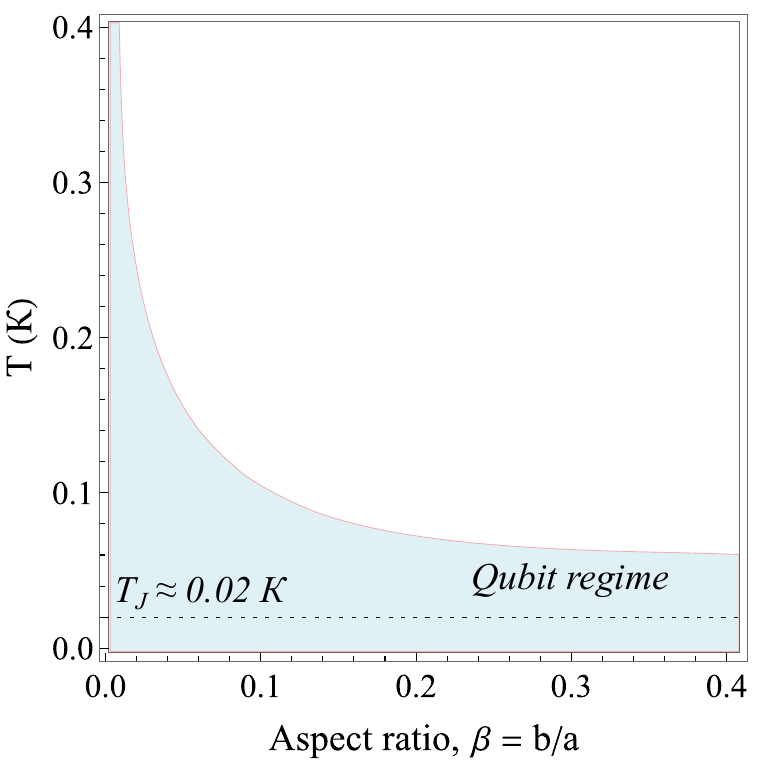}}
\caption{Diagram of the realization qubit regime in coordinates temperature -- aspect ratio $\beta = a/b$. The blue area in the diagram corresponds to the realization of the qubit regime; $T_{J}$ is characteristic temperature for realization of the superconducting Josephson qubit.  }
\label{Fig3}
\end{center}
\end{figure}

Taking into account all these limitations, we can identify a temperature range and characteristic sample dimensions suitable for the realization of a specific quantum regime. Based on inequality (\ref{eq23}), we construct a diagram in the coordinates temperature –- aspect ratio $\beta = b/a$, for fixed major semi-axes length $a = 10^{-6}$~cm and characteristic length $D = 0.05$ (corresponding to $d = 1$~nm), to demonstrate the parameter space where an MTJ-based qubit can be realized (see blue region in Fig. \ref{Fig3}).
Figure~\ref{Fig3} shows that the required temperatures lie within the range of $0$–$0.075$~K for high values of the aspect ratio. The permissible temperature increases as $\beta$ decreases. On the other hand, at fixed aspect ratio, the critical temperature is essentially independent of $D$. However, values of $\beta \leq 0.1$ are difficult to achieve in practice.
As a result, changes in the geometric dimensions of the sample have a negligible effect on the critical temperature, which can be estimated as $T \sim \hbar\omega_r/k \approx 360$~mK for ferromagnetic films under optimal conditions ($\omega_r \approx 0.47 \cdot 10^{11}$~rad/s). It is worth noting that this limit can be raised by increasing the magnon resonance energy. Specifically, for ferrimagnetic or antiferromagnetic materials, the resonance frequency can reach $\sim 1$~THz~\cite{10.1063/1.4958855,PhysRevLett.117.087203}. Therefore, using ferrimagnetic or antiferromagnetic films as the free layer allows the temperature restriction to be relaxed to approximately $0.5$–$1$~K~\cite{PhysRevB.95.144402}.
\par

Here, we provide the final numerical comparison with Josephson junction theory to complete the analysis of the isomorphism between the two systems. In the case of superconducting Josephson junctions, the key parameters are the dimensionless dissipation constant $\alpha_J$ and the plasma frequency $\omega_J = \sqrt{2eI_c/\hbar C}$, which was mentioned earlier in the text, where $R$ and $C$ are the resistance and capacitance of the junction, and $I_c$ is the critical current. Without considering the specific types of qubits, typical values of these quantities are $\omega_J \sim 6$~GHz and $\alpha_J \sim 10^{-3} - 10^{-5}$~\cite{PhysRevLett.97.050502,BLACKBURN20102827,BLACKBURN20161}. 
The corresponding parameters for MTJs — namely, the Gilbert damping constant $\alpha$ and the resonance frequency $\omega_r$ — are of the same order of magnitude ~\cite{PhysRevLett.107.066604,10.1038/s41467-018-05732-1}. However, $\omega_r$ in MTJs may be lower than the plasma frequency in Josephson junctions. One of the most important characteristics in superconducting qubits is the coherence time, which reflects the distinguishability of quantum states and is directly linked to the dissipation parameter $\alpha_J$. In Josephson-based systems, coherence times can reach up to 100~$\mu$s. Given the similarity in parameter values, it is reasonable to expect comparable coherence times for MTJ-based qubits, although they may be somewhat shorter. Notably, the Gilbert damping parameter in MTJs can be dynamically controlled using spin currents, potentially enhancing coherence times. This issue will be discussed in the next section. The temperature requirements to observe quantum behavior are typically stricter in Josephson junctions, often limited to around 20~mK~\cite{BLACKBURN20161}. This characteristic transition temperature is shown as a black dotted line in Fig.~\ref{Fig3}. Nevertheless, under certain experimental conditions, quantum behavior has been observed at temperatures above 300~mK~\cite{PhysRevLett.93.107002}.
\par

Thus, despite the similarity in key parameters such as resonance/plasma frequency and dissipation, the two systems differ only in their temperature limitations. This distinction arises primarily from the possibility of employing ferrimagnetic or antiferromagnetic materials in MTJs. For ferromagnetic films, the critical temperatures are approximately equivalent to those observed in superconducting systems. A summary of the main comparison parameters for Josephson and MTJ-based qubits is provided in Table~\ref{Tab2}.
\par

By varying the sample volume and the external magnetic field, we identify three distinct types of MTJ-based qubits: transmon (when $E_1 \ll E_H^{\parallel}$ and $E_2 \ll E_H^{\parallel}$), charge (when $E_1 \gg E_2$ and $E_1 \gg E_H^{\parallel}$), and flux ($E_1 \ll E_H^{\parallel}$ and $E_H^{\parallel} \sim E_2$), following the established classification in superconducting systems~\cite{doi:10.1063/1.2780165}. The main characteristics of each regime are summarized in Table~\ref{Tab3}.
To demonstrate the transition between regimes, we consider a volume range of $V \sim 10^{-19}$ to $10^{-21}$~cm$^3$. Within this range, transmon and flux regimes are found for $V \geq 10^{-21}$~cm$^3$, whereas the charge qubit approximation becomes valid for $V \leq 10^{-22}$~cm$^3$. All numerical values in Table~\ref{Tab3} are calculated for $N = 100$, $H^{\parallel} = 100$~Oe, $\beta = 0.4$, and $a = 10^{-6}$~cm. Note that the data in the Tables \ref{Tab1} and \ref{Tab3} correlate by the ratio of characteristic values, for example, $E_1/E_2$. This result is a consequence of the fixation of the number of spins $N = 100$. The characteristic values for $\omega_1$ and $\omega_2$ are $\sim 1 \cdot 10^9$ rad/s and $0.1 \cdot 10^9$ rad/s, respectively. These insights form the basis for the construction of analytical models applicable to each qubit type. 

\subsection{Transmon MTJ-based qubit}

\begin{figure}[h!]
\begin{center}
\center{\includegraphics[width=1.01\linewidth]{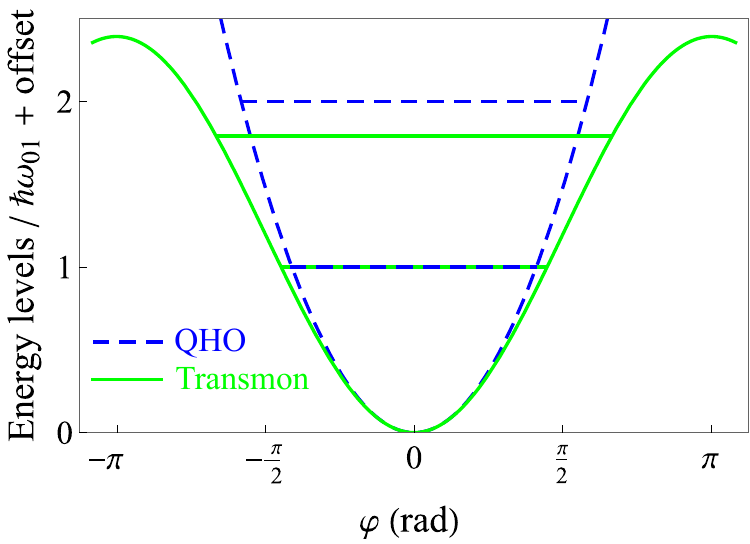}}
\caption{Schematic illustration of quantized energy levels: blue dotted curve corresponds to a quantum harmonic oscillator (QHO), while the green solid curve represents the anharmonic potential and nonequidistant energy levels of a transmon qubit.}
\label{Fig4}
\end{center}
\end{figure}

We consider the transmon MTJ-based qubit under the conditions $E_1 \ll E_2$, $E_1 \ll E^{\parallel}_H$, and $E_2 \ll E^{\parallel}_H$. In this regime, all energy levels are localized near the minimum of the effective potential, corresponding to $\varphi = 0$. We examine the stationary case $j(t) = 0$ and write the Hamiltonian of the transmon MTJ-based qubit using Eq.~(\ref{eq22}) in the following form (we assume $m^{\perp}_H = 0$, consistent with the transmon regime):
\begin{equation}\label{eq8}
\begin{gathered}
\mathcal{\hat{H}}_0 = 4E_1\hat{m}^2_z - E^{\parallel}_H\cos \hat{\varphi}
\end{gathered}. 
\end{equation}
Due to the concentration of energy levels near the minimum of the potential energy landscape, we expand the Hamiltonian around the equilibrium point $\varphi = 0$. Following the standard approach in quantum harmonic oscillator theory, we introduce the annihilation and creation operators:
 \begin{equation}\label{eq9}
\begin{gathered}
\hat{m}_z = -i\Big(\dfrac{E^{\parallel}_H}{32E_1}\Big)^{1/4}(\hat{c} - \hat{c}^\dagger), \; \hat{\varphi} = \Big(\dfrac{2E_1}{E^{\parallel}_H}\Big)^{1/4}(\hat{c} + \hat{c}^\dagger)
\end{gathered}. 
\end{equation}
Here, we define the annihilation operator for the MTJ-based transmon qubit as $\hat{c} = \sum_m \sqrt{m+1} \ket{m}\bra{m+1}$. Substituting the operator representation (\ref{eq9}) into the Hamiltonian expression (\ref{eq8}), we obtain the Hamiltonian of the MTJ-based transmon qubit:
\begin{equation}\label{eq10}
\begin{gathered}
\mathcal{\hat{H}}_0 = \sqrt{8 E_1 E^{\parallel}_H}\hat{c}^\dagger\hat{c} - \dfrac{E_1}{12}(\hat{c}^\dagger + \hat{c})^4 
\end{gathered}.
\end{equation}
Equation (\ref{eq10}) corresponds to the transmon Hamiltonian in Josephson junction theory~\cite{doi:10.1063/1.2780165}. We define the transition frequency between the two lowest energy levels and the anharmonicity as $\omega_{01} = \sqrt{8E_1 E^{\parallel}_H}/\hbar$ and $\delta = -E_1$, respectively. Using these definitions and the annihilation operator formalism, the Hamiltonian (\ref{eq10}) can be rewritten in the following form:
\begin{equation}\label{eq11}
\begin{gathered}
\mathcal{\hat{H}}_0 = \hbar\omega_{01}\hat{c}^\dagger\hat{c} + \dfrac{\delta}{2}((\hat{c}^\dagger\hat{c})^2 + \hat{c}^\dagger\hat{c}) =\\
= \sum_m\Big(\Big(\hbar\omega_{01} - \dfrac{\delta}{2}\Big)m +\dfrac{\delta}{2} m^2\Big)\ket{m}\bra{m} = 
\\ \sum_m\varepsilon^{tr}_m\ket{m}\bra{m},
\end{gathered}
\end{equation}
where $\varepsilon^{\mathrm{tr}}_m = \left(\hbar\omega_{01} - \frac{\delta}{2}\right)m + \frac{\delta}{2} m^2$ are the energy levels of the transmon, normalized with respect to the zero energy level. The potential landscape and quantized energy levels for the transmon, in comparison to those of a quantum harmonic oscillator, are illustrated in Fig.~\ref{Fig4}.
\par

Let us determine the important parameters of this qubit type. These include the transition frequency between the two lowest energy levels $\omega_{01}$, the critical temperature $T$, and the magnitude of the anharmonicity $|\delta/\hbar\omega_{01}|$, which determines the distinguishability between energy transitions. To quantify these, we consider three different values of the number of spins in the magnetic volume: $N = 10$, $N = 50$, and $N = 100$. The following fixed parameters are used for the calculations: $V = 2.51 \cdot 10^{-19}$ cm$^3$, $d = 4$ nm, $\beta  = 0.4$, $a = 10^{-6}$ cm, $S = 125$ nm$^2$, and $H^{\parallel} = 100$ Oe.
All numerical results for the transmon MTJ-based qubit are summarized in Table \ref{Tab4}. From this, we conclude that the transmon regime can be realized using experimentally feasible geometric parameters. Increasing the spin density $N/V$ results in a higher critical temperature $T$, while still remaining within the expected low-temperature range. As a result, a two-level MTJ-based qubit with realistic characteristics and described by the Hamiltonian (\ref{eq10}) is achievable.
However, the distinguishability of the energy levels is critically dependent on the degree of anharmonicity $|\delta/\hbar\omega_{01}|$. If this parameter is too small, the energy spectrum becomes nearly harmonic, making it difficult to isolate a true two-level system. This may present a challenge for practical realization of the qubit regime in such cases.
\begin{table}[!]
\caption{\label{Tab4}Main characteristics of the transmon MTJ based qubit for $V = 2.51 \cdot 10^{-19}$ cm$^3$, $d = 4$ nm, $\beta  = 0.4$, $a = 10^{-6}$ cm , $S = 125$ nm$^2$, $H^{\parallel} = 100$ Oe  }
\begin{ruledtabular}
\begin{tabular}{c|ccc}
\small \centering
\renewcommand{\tabcolsep}{0.14cm}
\rule{0mm}{12pt}
Spins, $N$ & $N = 10$ & $N = 50$ & $N = 100$ \\ \hline
\rule{0mm}{12pt}
$\hbar\omega_{01}$, erg & $4.38 \cdot 10^{-18}$ & $4.89\cdot10^{-17}$ & $1.38\cdot 10^{-16}$  \\
\rule{0mm}{12pt}
$\omega_{01}$, rad/s & $4.15 \cdot 10^{9}$ & $4.64\cdot10^{10}$ & $1.31\cdot 10^{11}$  \\
\rule{0mm}{12pt}
$T$, mK & 31 & 354 & 1003  \\
\rule{0mm}{12pt}
$|\delta/\hbar\omega_{01}|$ & 0.03 & 0.06 & 0.09  \\
\end{tabular}
\end{ruledtabular}
\end{table}
\par

After fully describing the transmon realization, we now turn to the non-stationary regime, where $j(t) \neq 0$, in order to obtain the time-dependent qubit Hamiltonian. By using equations (\ref{eq11}) and (\ref{eq10}), the Hamiltonian can be expressed in the following form:

\begin{equation}\label{eq12}
\begin{gathered}
\mathcal{\hat{H}} = \hbar\omega_{01}\hat{c}^\dagger\hat{c} + \dfrac{\delta}{2}((\hat{c}^\dagger\hat{c})^2 + \hat{c}^\dagger\hat{c}) - \overline{W}(t)(\hat{c} + \hat{c}^{\dagger})
\end{gathered},
\end{equation}
where the last term describes the influence of the spin current, and $\overline{W}(t) = W(t)\left(2E_1/E^{\parallel}_H\right)^{1/4}$. Typically, the electric current can be decomposed into two components: a constant displacement current and a time-dependent current. Therefore, we write $W(t) = W_{\mathrm{dis}} + \widetilde{W}(t)$, where $W_{\mathrm{dis}}$ represents the spin-current energy associated with the displacement current, and $\widetilde{W}(t)$ denotes the time-dependent component.
\par
Let us now derive the Hamiltonian of the transmon in the presence of an electric current from equation~(\ref{eq12}). We treat the transmon as a qubit, considering only the first two energy levels. By applying equation~(\ref{eq11}) and shifting the energy levels by a constant, we obtain:
\begin{equation}\label{eq13}
\begin{gathered}
\mathcal{\hat{H}}_0 = -\dfrac{1}{2}\hbar\omega_{01}\ket{0}\bra{0} + \dfrac{1}{2}\hbar\omega_{01}\ket{1}\bra{1} = \\ 
= -\dfrac{1}{2}\hbar\omega_{01}\hat{\sigma}_z,
\end{gathered}
\end{equation}
where $\hat{\sigma}_z$ is the Pauli matrix. Considering the transmon as a qubit, we can introduce the raising and lowering operators $\hat{\sigma}^{\pm} = \frac{1}{2}(\hat{\sigma}_x \mp i \hat{\sigma}_y)$, which act on the basis states as $\hat{\sigma}^{+}\ket{0} = \ket{1}$ and $\hat{\sigma}^{-}\ket{1} = \ket{0}$. It is worth noting that the action of these operators resembles that of the annihilation and creation operators introduced earlier for the lowest energy levels. In this case, it is straightforward to derive the displacement part of the Hamiltonian, and according to equations~(\ref{eq12}) and~(\ref{eq13}), we write:
\begin{equation}\label{eq14}
\begin{gathered}
\mathcal{\hat{H}} = \mathcal{\hat{H}}_{0} - W_{dis}(\hat{\sigma}^{+} + \hat{\sigma}^{-}) = \\ = -\dfrac{1}{2}\hbar\omega_{01}\hat{\sigma}_z - W_{dis}\hat{\sigma}_x,
\end{gathered}
\end{equation}

where the displacement current appears as an $x$-rotation on the Bloch sphere. The time-dependent part of the Hamiltonian, $\widetilde{W}(t)$, enables the calculation of the transition probability to higher energy levels. Including this time-dependent perturbation is analogous to the observation of Rabi oscillations in superconducting Josephson junctions.
\par
The final aspect to address is the determination of the coherence time. To this end, we introduce a term $\hat{H}_{\alpha} = M_s V h_{\mathrm{eff}}(t)\hat{\varphi}$ into the Hamiltonian (\ref{eq8}), which accounts for thermal, quantum, and other types of fluctuations. Initially, we also assume that $j(t) = 0$. This term is equivalent to introducing an effective magnetic field directed along the $y$-axis into the original dynamical equation (\ref{eq4}). Such an introduction is justified since the system is considered near its equilibrium state ($E^{\parallel}_H \gg E_1$ and $\varphi_0 \approx 0$), allowing us to treat the response to external perturbations linearly.
According to the fluctuation-dissipation theorem, the correlation function for this effective field is given by $\langle h_{\mathrm{eff}}(t) h_{\mathrm{eff}}(0) \rangle = \int \exp(-i\omega t) \chi(\omega)\, d\omega/2\pi$, where the spectral density is $\chi(\omega) = \frac{\alpha \hbar \omega}{2 V M_s \gamma} \coth{(\hbar \omega / 2kT)}$ \cite{PhysRevB.95.144402,PhysRevB.67.094510,landau2013statistical}. Following the procedure described in \cite{PhysRevB.95.144402}, we obtain the expression for the coherence time as $1/T_1 = 2 E_1 M_s V \alpha/(\hbar^2 \gamma) \sim \alpha \omega_1$, yielding a characteristic value of $T_1 \sim 1\,\mu\text{s}$. This estimate is comparable to that observed in Josephson junction systems.
\par
It is also important to clarify the role of Gilbert damping. The presence of a spin current modifies the damping parameter $\alpha$ in the LLG equation, leading to an effective form $\alpha \rightarrow \alpha_{\mathrm{eff}}$. Specifically, when $j(t) \neq 0$ the spin current reduces the effective damping, and the dissipation can be described by $\alpha_{\mathrm{eff}} = |\alpha - 2j/\omega_1|$. This approximation is derived by eliminating $d\mathbf{m}/dt$ from the right-hand side of equation (\ref{eq1}) and rewriting it in terms of effective fields. Importantly, the compensation of the damping term is only possible if the polarization vector $\mathbf{m}_{\mathrm{ref}}$ lies within the film plane or has a non-zero projection on it (e.g., along the $x$-axis). In such case, the general form of the Hamiltonian will change and the time-dependent component will take the following form:
$\mathcal{H}_T(t) = \Big(-\dfrac{j^2(t)M_sV}{4\gamma\omega_1}\cos2\varphi + \dfrac{j(t)M_sV\omega^{\perp}_H}{\gamma\omega_1}\sin\varphi\Big)n_{\parallel} + \Big(- \dfrac{j(t)M_sV}{\gamma}\varphi\Big)n_{\perp}$.
The term preceding $n_{\parallel}$ corresponds to the in-plane component of the spin current, while the term before $n_{\perp}$ represents its perpendicular component. The angle between these components is given by $\tan\Theta = n_{\perp}/n_{\parallel}$. When a non-zero spin current is present, the correlation function of the effective field is modified due to the contribution of stochastic fluctuations arising from the spin current itself, which can become particularly significant at low temperatures \cite{PhysRevLett.101.066601}. Although the ability to control dissipation through spin currents provides a potential means to suppress these effects, a more detailed analysis is necessary to fully quantify this behavior. Nonetheless, the presented qubit architecture offers the advantage of tunable coherence time through engineered dissipation, which is a promising feature for quantum information applications.

\subsection{Flux MTJ-based qubit}
\begin{figure}[h!]
\begin{center}
\center{\includegraphics[width=1.01\linewidth]{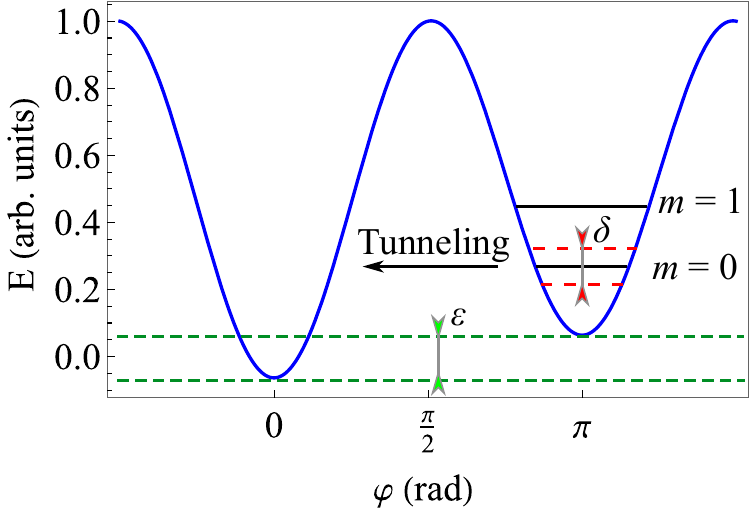}}
\caption{Schematical potential landscape and energy levels of the two well potential; the dotted lines demonstrate the hybridization of the quantum state due to tunneling.}
\label{Fig5}
\end{center}
\end{figure}
A flux MTJ-based qubit is realized under transmon-like conditions, with non-zero in-plane magnetic anisotropy and the presence of an in-plane magnetic field, under the conditions $E_1 \ll E^{\parallel}_H$ and $E^{\parallel}_H \sim E_2$. 
In the stationary case ($j(t) = 0$), the system exhibits a $2\pi$-periodic double-well potential, with ground states localized near the minima at $\varphi^{(1)} = 0$ and $\varphi^{(2)} = \pi$, as illustrated in Fig. \ref{Fig5}. 
This potential landscape emerges only when both anisotropy and Zeeman energy contributions are taken into account.
\par

Let us write the Hamiltonian for this structure using Eq.~(\ref{eq22}) (here we suppose that $m^{\perp}_H = 0$ due to flux regime):
\begin{equation}\label{eq15}
\begin{gathered}
\mathcal{\hat{H}}_0 = 4E_1\hat{m}_z^2 - E_2\cos2\hat{\varphi}  - E^{\parallel}_{H}\cos\hat{\varphi}
\end{gathered}.
\end{equation}
To build the qubit Hamiltonian for this case, we introduce the creation and annihilation operators similar to the transmon case. Thus, in the zeroth approximation, where the wells are separated from each other by an infinite potential barrier, these operators can be used to construct quantum states in each well. As a result, the zero energy wave function of the (\ref{eq15}) can be written as $\psi_0(\varphi) = \dfrac{1}{\pi^{1/4}\sqrt{\sigma_{\varphi}}}\exp{-\frac{(\varphi - \varphi^{(1,2)})^2}{2\sigma_{\varphi}^2}}$, where $\sigma_{\varphi} = \sqrt{\hbar/V m^*\omega_{01}}$ is the length of the wave function localization \cite{PhysRevB.59.2070}, $\omega_{01} = \sqrt{8(4E_2 + E^{\parallel}_{H})E_1}/\hbar$ can be found similar to the transmon case. The possibility of tunneling from the potential barrier with height $\Delta E = 2E_2 + E^{\parallel}_{H}$ leads to hybridization of the discrete energy levels, which are localized near the ground states. As a result, the energy splitting occurs and the two-level system is realized. The schematic of this process is shown in Fig. \ref{Fig5}. Using the instant approach \cite{coleman_1985} we can find the energy gap between the hybridized levels in action terms $S_{int}$ as $\delta = 4\hbar\omega_{01}\sqrt{\dfrac{S_{int}}{2\pi\hbar}}\exp{-S_{int}/\hbar}$, where $S_{int} \approx 4\Delta E/\omega_{01}$. Thus, for the symmetric double-well potential, the Hamiltonian can be written as $\hat{\mathcal{H}} = -\dfrac{1}{2}\delta\hat{\sigma}_x$. However, taking into account the difference between the bottoms of the potential energy $\varepsilon = 2 E^{\parallel}_H$, Hamiltonian of the flux MTJ-based qubit can be written as
\begin{equation}\label{eq16}
\begin{gathered}
\mathcal{\hat{H}}_0 = -\dfrac{1}{2}(\varepsilon\hat{\sigma}_z + \delta\hat{\sigma}_x)
\end{gathered}.
\end{equation}
\par
Note that the data findings from the transmon chapter and Table \ref{Tab4} are suitable for the flux qubit. All results for coherence time and external influence of the spin current are also valid for flux MTJ-based qubit. Therefore, we build analogues of the flux qubit from Josephson's transition theory with the Hamilonian (\ref{eq16}). Taking into account the main numerical results, we can conclude that flux and transmon MTJ-based qubits are the most advantageous for physical implementation. 

\subsection{Charge MTJ-based qubit}
\begin{figure}[h!]
\begin{center}
\center{\includegraphics[width=1\linewidth]{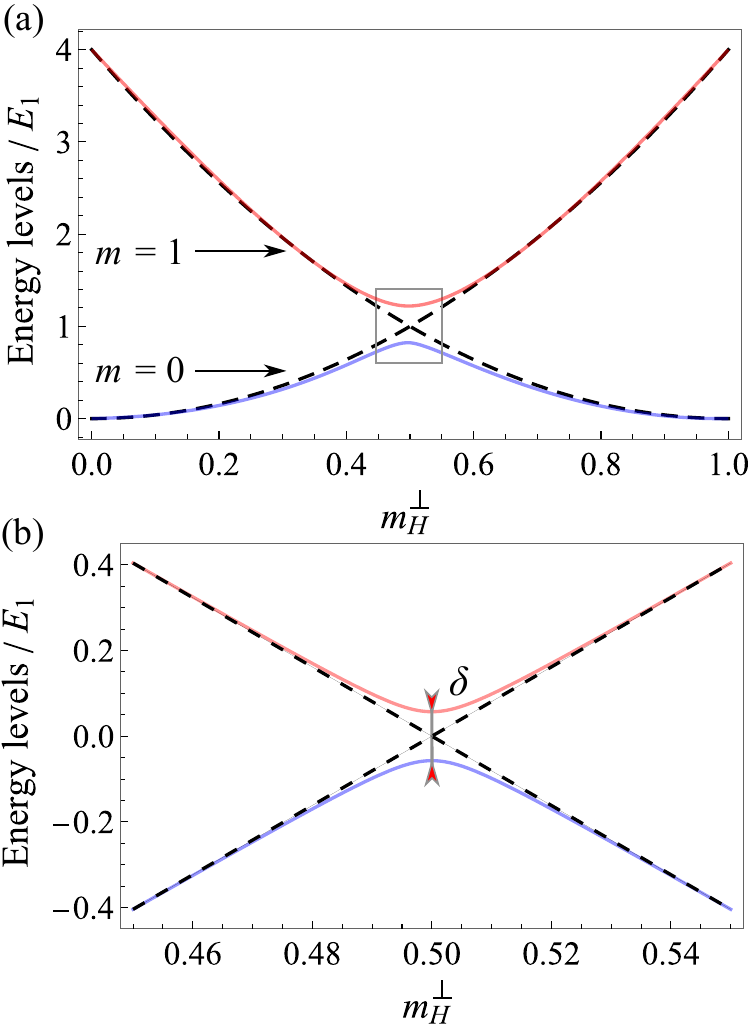}}
\caption{a) Schematic shape of the quantized energy levels of the charge MTJ-based qubit (solid lines). b) Zoomed area of the degeneration point of the two lower quantum states. }
\label{Fig6}
\end{center}
\end{figure}

Taking into account the data from Table \ref{Tab3}, we can conclude that the charge MTJ-based qubit is challenging to implement in practice, as it would require a layer thickness of approximately $d \approx 0.1$ nm. However, it should be noted that the form of the magnetic crystalline anisotropy constants may be modified due to the presence of an easy axis within the plane of the film or due to intrinsic rhombic anisotropy of the magnetic film, which is typical for orthoferrites. In such cases, the relationship between these parameters and the geometry of the magnetic film becomes significantly more complex. This observation suggests that, although this type of qubit is difficult to realize in ferromagnetic materials, it may still be feasible in antiferromagnetic or ferrimagnetic films. It is also worth noting that, in the zeroth-order approximation, the results for these films will coincide with those discussed below, up to numerical prefactors.
\par

The analog of the superconducting Josephson charge qubit is realized in the MTJ structure when $E_1 \gg E^{\parallel}_{H} \gg E_2$. In this regime, quantum states emerge in the absence of significant magnetostatic anisotropy. Assuming the stationary approximation $j(t) = 0$, the Hamiltonian of the MTJ according to Eq~(\ref{eq22}) takes the following form:
 \begin{equation}\label{eq17}
\begin{gathered}
\mathcal{\hat{H}}_0 = 4E_1(\hat{m}_z + m_H^{\perp})^2 - E^{\parallel}_H\cos\hat{\varphi}
\end{gathered}. 
\end{equation}

According to the charge-based assumption, we can neglect the Zeeman term in the Hamiltonian (\ref{eq17}) and obtain a qubit with integer spin projections along the $z$-axis, denoted by $m$. By introducing the operator definitions $\hat{m}_z\ket{m} = m\ket{m}$, $[\hat{\varphi}, \hat{m}_z] = i$, and $\exp{\left(\pm i\hat{\varphi}\right)}\ket{m} = \ket{m \pm 1}$ \cite{Kockum2019}, we can formulate the eigenvalue problem for the Hamiltonian (\ref{eq17}) as follows:
\begin{equation}\label{eq18}
\begin{gathered}
4E_1(\hat{m}_z - m^{\perp}_H)^2\ket{m} = \varepsilon^{ch}_m\ket{m}
\end{gathered},
\end{equation}

where $\ket{m}$ is the eigenvector of the reduced Hamiltonian (\ref{eq17}). The corresponding eigenvalue is given by $\varepsilon^{ch}_m = 4 E_1 (m - m^{\perp}_H)^2$. The energy of the state increases with the magnitude of the external magnetic field, as well as with the spin projection $m_z$, as illustrated in Fig.~\ref{Fig6}(a). 
However, at the specific value $m^{\perp}_H = 1/2$, the quantum states $\ket{0}$ and $\ket{1}$ become degenerate (see Fig.~\ref{Fig6}(a) and (b)). Introducing the Zeeman term $E^{\parallel}_{H}$ removes this degeneracy, resulting in a two-level system. The Hamiltonian can then be rewritten using mixed states \cite{doi:10.1063/1.2780165}:
\begin{equation}\label{eq19}
\begin{gathered}
\mathcal{\hat{H}}_0 = \sum_m 4 E_{1}(\hat{m}_z - m^{\perp}_H )^2\ket{m}\bra{m} - \\ - \dfrac{E^{\parallel}_H}{2}( \ket{m}\bra{m+1} + \ket{m+1}\bra{m}),
\end{gathered}  
\end{equation}
The last term can be obtained by expanding $\cos\hat{\varphi} = (e^{i\hat{\varphi}} + e^{-i\hat{\varphi}})/2$ in the basis $\ket{m}$. After shifting the energy levels by a constant and restricting consideration to only the first two energy levels in (\ref{eq19}), the Hamiltonian takes a form identical to that of the charge qubit in the Josephson junction theory \cite{doi:10.1063/1.2780165}.

\begin{equation}\label{eq20}
\begin{gathered}
\mathcal{\hat{H}}_0 = -\dfrac{1}{2}(\varepsilon\hat{\sigma}_z + \delta\hat{\sigma}_x),
\end{gathered}  
\end{equation}

where $\hat{\sigma}_{z,x}$ are the Pauli matrices, $\varepsilon = 4 E_1 (1 - 2 m^{\perp}_H)$, and $\delta = E^{\parallel}_H$. The eigenvalues of equation (\ref{eq20}) are $E_{\pm} = \pm \sqrt{(4E_1(1 - 2m^{\perp}_H))^2 + (E^{\parallel}_{H})^2}$, and the energy gap between these levels is approximately $2E^{\parallel}_{H}$ near $m^{\perp}_H = 1/2$.
\par
\begin{table}[h!]
\caption{\label{Tab5}Main characteristics of the charge MTJ based qubit for $V = 6.28\cdot10^{-21}$ cm$^3$, $d  = 0.1$ nm, $\beta  = 0.4$, $a = 10^{-6}$ cm, $S = 125$ nm$^2$. }
\small \centering
\begin{ruledtabular}
\begin{tabular}{c|ccc}
\small \centering
\renewcommand{\tabcolsep}{0.14cm}
\rule{0mm}{12pt}
Spins, $N$ & $N = 1$ & $N = 10$ & $N = 100$ \\ \hline
\rule{0mm}{12pt}
$H^{\parallel
}$, Oe & $5$ & $10$ & $100$  \\
\rule{0mm}{12pt}
$\hbar\omega_{01}$, erg & $3.33 \cdot 10^{-19}$& $3.33\cdot10^{-17}$ & $3.33\cdot 10^{-15}$  \\
\rule{0mm}{12pt}
$\omega_{01}$, rad/s & $3.16 \cdot 10^{8}$ & $3.16 \cdot 10^{10}$ & $3.16\cdot 10^{12}$  \\
\rule{0mm}{12pt}
$T$, mK & 2 & 241 & 2415  \\
\end{tabular}
\end{ruledtabular}
\end{table}

By analogy with the previous subsections, we calculate the key parameters for a charge MTJ-based qubit. The following parameters were used for the calculations: $V = 6.28\cdot10^{-21}$ cm$^3$, $\beta = 0.4$, $a = 10^{-6}$ cm, and $S = 125$ nm$^2$. The numerical data corresponding to the transition into the quantum regime are summarized in Table \ref{Tab5}. 
The results show that the transition to a charge qubit state is feasible under the condition $N = 10$, with $\hbar\omega_{01} \approx 3.33\cdot10^{-17}$ erg and $\omega_{01} \approx 3.16 \cdot 10^{8}$ rad/s, yielding a critical temperature of $T = 241$ mK. 
Thus, a numerical model of a charge qubit based on a magnetic tunnel junction has been established.

\section{Conclusion}

In this work, we have presented a comprehensive theoretical analysis of a Josephson-like magnetic tunnel junction (MTJ) structure capable of exhibiting macroscopic quantum spin dynamics. By formulating the magnetization dynamics in the low-dissipation regime, we have constructed a Lagrangian and Hamiltonian framework that is formally isomorphic to the theory of superconducting Josephson junction qubits.
\par
We have demonstrated that the azimuthal angle $\varphi$ and the projection of the magnetization onto the $z$-axis $M_z$ in the MTJ play roles analogous to the superconducting phase and charge operators, respectively. Based on this analogy, we derived the quantum Hamiltonian of the MTJ and explored its applicability to different qubit regimes — namely, transmon-, flux-, and charge-like spintronic qubits — depending on the energy scale hierarchy between the effective kinetic energy $E_1$, the magnetic anisotropy energy $E_2$, and the Zeeman energy $E_H^{\parallel}$.
\par
We identified the geometric, material, and field conditions required to achieve quantum coherence in MTJs, including constraints on anisotropy ratios, Gilbert damping $\alpha$, resonant frequency $\omega_r$, and temperature $T$. Special attention was given to the transmon regime, where we derived a two-level system Hamiltonian and showed how spin currents can be used to tune system parameters and influence coherence.
\par
Furthermore, we discussed the problem of coherence time $T_1$ in the presence of dissipation and noise, showing that the effective damping parameter $\alpha_{\mathrm{eff}}$ can be tuned via spin current injection, thus offering a mechanism for dynamical control of decoherence. Our analysis indicates that coherence times in the order of microseconds may be achievable under suitable conditions, comparable to those in superconducting qubits.
\par
Numerical estimates suggest that transmon and flux MTJ-based qubits are the most promising for experimental implementation due to their relaxed requirements on sample dimensions and damping, whereas the realization of a charge-type MTJ qubit appears less feasible.
\par
Altogether, our findings establish a theoretical foundation for MTJ-based quantum devices and suggest a pathway toward realizing fully spintronic, CMOS-compatible two-level quantum systems. Such spintronic qubits could open a new avenue in quantum information processing by integrating memory, logic, and quantum coherence in a unified nanoscale platform.

This work was supported by the RSF grant No. 25-12-00302.

\providecommand{\noopsort}[1]{}\providecommand{\singleletter}[1]{#1}%


\begin{thebibliography}{0}%
\makeatletter
\providecommand \@ifxundefined [1]{%
 \@ifx{#1\undefined}
}%
\providecommand \@ifnum [1]{%
 \ifnum #1\expandafter \@firstoftwo
 \else \expandafter \@secondoftwo
 \fi
}%
\providecommand \@ifx [1]{%
 \ifx #1\expandafter \@firstoftwo
 \else \expandafter \@secondoftwo
 \fi
}%
\providecommand \natexlab [1]{#1}%
\providecommand \enquote  [1]{``#1''}%
\providecommand \bibnamefont  [1]{#1}%
\providecommand \bibfnamefont [1]{#1}%
\providecommand \citenamefont [1]{#1}%
\providecommand \href@noop [0]{\@secondoftwo}%
\providecommand \href [0]{\begingroup \@sanitize@url \@href}%
\providecommand \@href[1]{\@@startlink{#1}\@@href}%
\providecommand \@@href[1]{\endgroup#1\@@endlink}%
\providecommand \@sanitize@url [0]{\catcode `\\12\catcode `\$12\catcode
  `\&12\catcode `\#12\catcode `\^12\catcode `\_12\catcode `\%12\relax}%
\providecommand \@@startlink[1]{}%
\providecommand \@@endlink[0]{}%
\providecommand \url  [0]{\begingroup\@sanitize@url \@url }%
\providecommand \@url [1]{\endgroup\@href {#1}{\urlprefix }}%
\providecommand \urlprefix  [0]{URL }%
\providecommand \Eprint [0]{\href }%
\providecommand \doibase [0]{https://doi.org/}%
\providecommand \selectlanguage [0]{\@gobble}%
\providecommand \bibinfo  [0]{\@secondoftwo}%
\providecommand \bibfield  [0]{\@secondoftwo}%
\providecommand \translation [1]{[#1]}%
\providecommand \BibitemOpen [0]{}%
\providecommand \bibitemStop [0]{}%
\providecommand \bibitemNoStop [0]{.\EOS\space}%
\providecommand \EOS [0]{\spacefactor3000\relax}%
\providecommand \BibitemShut  [1]{\csname bibitem#1\endcsname}%
\let\auto@bib@innerbib\@empty
\end{thebibliography}%


\begin{thebibliography}{10}

\bibitem{https://doi.org/10.1002/1521-3978(200009)48:9/11<771::AID-PROP771>3.0.CO;2-E}
David~P. DiVincenzo.
\newblock The physical implementation of quantum computation.
\newblock {\em Fortschritte der Physik}, 48(9-11):771--783, 2000.

\bibitem{doi:10.1098/rsta.2003.1227}
A.~G.~J. MacFarlane, Jonathan~P. Dowling, and Gerard~J. Milburn.
\newblock Quantum technology: the second quantum revolution.
\newblock {\em Philosophical Transactions of the Royal Society of London.
  Series A: Mathematical, Physical and Engineering Sciences},
  361(1809):1655--1674, 2003.

\bibitem{doi:10.1126/science.273.5278.1073}
Seth Lloyd.
\newblock Universal quantum simulators.
\newblock {\em Science}, 273(5278):1073--1078, 1996.

\bibitem{doi:10.1063/1.2780165}
G.~Wendin and V.~S. Shumeiko.
\newblock Quantum bits with josephson junctions (review article).
\newblock {\em Low Temperature Physics}, 33(9):724--744, 2007.

\bibitem{10.1063/1.5089550}
P.~Krantz, M.~Kjaergaard, F.~Yan, T.~P. Orlando, S.~Gustavsson, and W.~D.
  Oliver.
\newblock {A quantum engineer's guide to superconducting qubits}.
\newblock {\em Applied Physics Reviews}, 6(2):021318, 06 2019.

\bibitem{10.1063/1.2155757}
J.~Q. You and Franco Nori.
\newblock {Superconducting Circuits and Quantum Information}.
\newblock {\em Physics Today}, 58(11):42--47, 11 2005.

\bibitem{Zvezdin2002}
A.~K. Zvezdin and K.~A. Zvezdin.
\newblock Spin-current-induced classical and quantum effects in the dynamics of
  a mesoscopic magnet.
\newblock {\em Journal of Experimental and Theoretical Physics}, 95:762--767,
  2002.

\bibitem{PhysRevLett.127.067201}
Christina Psaroudaki and Christos Panagopoulos.
\newblock Skyrmion qubits: A new class of quantum logic elements based on
  nanoscale magnetization.
\newblock {\em Phys. Rev. Lett.}, 127:067201, Aug 2021.

\bibitem{PhysRevResearch.5.033166}
Ji~Zou, Stefano Bosco, Banabir Pal, Stuart S.~P. Parkin, Jelena Klinovaja, and
  Daniel Loss.
\newblock Quantum computing on magnetic racetracks with flying domain wall
  qubits.
\newblock {\em Phys. Rev. Res.}, 5:033166, Sep 2023.

\bibitem{Demokritov2006}
S.~O. Demokritov, V.~E. Demidov, O.~Dzyapko, G.~A. Melkov, A.~A. Serga, and
  A.~N. Slavin.
\newblock Bose–einstein condensation of quasi-equilibrium magnons at room
  temperature under pumping.
\newblock {\em Nature}, 443, 2006.

\bibitem{PhysRevLett.102.187205}
A.~V. Chumak, G.~A. Melkov, V.~E. Demidov, O.~Dzyapko, V.~L. Safonov, and S.~O.
  Demokritov.
\newblock Bose-einstein condensation of magnons under incoherent pumping.
\newblock {\em Phys. Rev. Lett.}, 102:187205, May 2009.

\bibitem{Bozhko2016}
Dmytro~A. Bozhko, Alexander~A. Serga, Peter Clausen, Vitaliy~I. Vasyuchka,
  Frank Heussner, Gennadii~A. Melkov, Anna Pomyalov, Victor~S. L’vov, and
  Burkard Hillebrands.
\newblock Supercurrent in a room-temperature bose–einstein magnon condensate.
\newblock {\em Nature Physics}, 12:1057--1062, 2016.

\bibitem{Divinskiy2021}
B.~Divinskiy, H.~Merbouche, V.~E. Demidov, K.~O. Nikolaev, L.~Soumah,
  D.~Gouéré, R.~Lebrun, V.~Cros, Jamal~Ben Youssef, P.~Bortolotti, A.~Anane,
  and S.~O. Demokritov.
\newblock Evidence for spin current driven bose-einstein condensation of
  magnons.
\newblock {\em Nature Communications}, 12, 2021.

\bibitem{Rüegg2003}
Ch. Rüegg, N.~Cavadini, A.~Furrer, H.-U. Güdel, K.~Krämer, H.~Mutka,
  A.~Wildes, K.~Habicht, and P.~Vorderwisch.
\newblock Bose–einstein condensation of the triplet states in the magnetic
  insulator tlcucl3.
\newblock {\em Nature}, 423:62--65, 2003.

\bibitem{vetoshko2020bose}
Petr~Mikhailovich Vetoshko, Grigorii~Alekseevich Knyazev, Aleksei~Nikolaevich
  Kuzmichev, AA~Kholin, Vladimir~Igorevich Belotelov, and Yu~M Bunkov.
\newblock Bose—einstein condensation and spin superfluidity of magnons in a
  perpendicularly magnetized yttrium iron garnet film.
\newblock {\em JETP Letters}, 112(5):299--304, 2020.

\bibitem{bunkov2020features}
Yu~M Bunkov and D~Konstantinov.
\newblock Features of the coupled nuclear--electron spin precession in the
  bose--einstein condensate of magnons.
\newblock {\em JETP Letters}, 112(2):95--100, 2020.

\bibitem{petrov2024transition}
Petr~Evgen'evich Petrov, Grigorii~Alekseevich Knyazev, Aleksei~Nikolaevich
  Kuzmichev, Petr~Mikhailovich Vetoshko, Vladimir~Igorevich Belotelov, and Yu~M
  Bunkov.
\newblock Transition to a magnon bose--einstein condensate.
\newblock {\em JETP Letters}, 119(2):118--122, 2024.

\bibitem{Khymyn2017}
Roman Khymyn, Ivan Lisenkov, Vasyl Tiberkevich, Boris~A. Ivanov, and Andrei
  Slavin.
\newblock Antiferromagnetic thz-frequency josephson-like oscillator driven by
  spin current.
\newblock {\em Scientific Reports}, 7(1):43705, 2017.

\bibitem{PhysRevB.85.014523}
A.~Moor, A.~F. Volkov, and K.~B. Efetov.
\newblock Josephson-like spin current in junctions composed of antiferromagnets
  and ferromagnets.
\newblock {\em Phys. Rev. B}, 85:014523, Jan 2012.

\bibitem{PhysRevB.94.094434}
Yizhou Liu, Gen Yin, Jiadong Zang, Roger~K. Lake, and Yafis Barlas.
\newblock Spin-josephson effects in exchange coupled antiferromagnetic
  insulators.
\newblock {\em Phys. Rev. B}, 94:094434, Sep 2016.

\bibitem{PhysRevB.104.104402}
Kouki Nakata.
\newblock Optomagnonic josephson effect in antiferromagnets.
\newblock {\em Phys. Rev. B}, 104:104402, Sep 2021.

\bibitem{PhysRevApplied.13.034035}
H.~J. Waring, N.~A.~B. Johansson, I.~J. Vera-Marun, and T.~Thomson.
\newblock Zero-field optic mode beyond 20 ghz in a synthetic antiferromagnet.
\newblock {\em Phys. Rev. Appl.}, 13:034035, Mar 2020.

\bibitem{Khvalkovskiy_2013}
A~V Khvalkovskiy, D~Apalkov, S~Watts, R~Chepulskii, R~S Beach, A~Ong, X~Tang,
  A~Driskill-Smith, W~H Butler, P~B Visscher, D~Lottis, E~Chen, V~Nikitin, and
  M~Krounbi.
\newblock Basic principles of stt-mram cell operation in memory arrays.
\newblock {\em Journal of Physics D: Applied Physics}, 46(7):074001, jan 2013.

\bibitem{Kiselev2003}
S.~I. Kiselev, J.~C. Sankey, I.~N. Krivorotov, N.~C. Emley, R.~J. Schoelkopf,
  R.~A. Buhrman, and D.~C. Ralph.
\newblock Microwave oscillations of a nanomagnet driven by a spin-polarized
  current.
\newblock {\em Nature}, 425(7):380--383, 2003.

\bibitem{PhysRevLett.92.027201}
W.~H. Rippard, M.~R. Pufall, S.~Kaka, S.~E. Russek, and T.~J. Silva.
\newblock Direct-current induced dynamics in
  ${\mathrm{c}\mathrm{o}}_{90}{\mathrm{f}\mathrm{e}}_{10}/{\mathrm{n}\mathrm{i}}_{80}{\mathrm{f}\mathrm{e}}_{20}$
  point contacts.
\newblock {\em Phys. Rev. Lett.}, 92:027201, Jan 2004.

\bibitem{PhysRevApplied.22.024019}
C.K. Safeer, Paul~S. Keatley, Witold Skowro\ifmmode~\acute{n}\else
  \'{n}\fi{}ski, Jakub Mojsiejuk, Kay Yakushiji, Akio Fukushima, Shinji Yuasa,
  Daniel Bedau, F\`elix Casanova, Luis~E. Hueso, Robert~J. Hicken, Daniele
  Pinna, Gerrit van~der Laan, and Thorsten Hesjedal.
\newblock Magnetization dynamics driven by displacement currents across a
  magnetic tunnel junction.
\newblock {\em Phys. Rev. Appl.}, 22:024019, Aug 2024.

\bibitem{10.1063/1.5044435}
Steven Louis, Olga Sulymenko, Vasil Tiberkevich, Jia Li, Daniel Aloi, Oleksandr
  Prokopenko, Ilya Krivorotov, Elena Bankowski, Thomas Meitzler, and Andrei
  Slavin.
\newblock Ultra-fast wide band spectrum analyzer based on a rapidly tuned
  spin-torque nano-oscillator.
\newblock {\em Applied Physics Letters}, 113(11):112401, 09 2018.

\bibitem{Romera2018}
Miguel Romera, Philippe Talatchian, Sumito Tsunegi, Flavio Abreu~Araujo,
  Vincent Cros, Paolo Bortolotti, Juan Trastoy, Kay Yakushiji, Akio Fukushima,
  Hitoshi Kubota, Shinji Yuasa, Maxence Ernoult, Damir Vodenicarevic, Tifenn
  Hirtzlin, Nicolas Locatelli, Damien Querlioz, and Julie Grollier.
\newblock Vowel recognition with four coupled spin-torque nano-oscillators.
\newblock {\em Nature}, 563:230--234, 10 2018.

\bibitem{Fukushima_2014}
Akio Fukushima, Takayuki Seki, Kay Yakushiji, Hitoshi Kubota, Hiroshi Imamura,
  Shinji Yuasa, and Koji Ando.
\newblock Spin dice: A scalable truly random number generator based on
  spintronics.
\newblock {\em Applied Physics Express}, 7(8):083001, jul 2014.

\bibitem{montoya2019magnetization}
Eric~Arturo Montoya, Salvatore Perna, Yu-Jin Chen, Jordan~A Katine,
  Massimiliano d’Aquino, Claudio Serpico, and Ilya~N Krivorotov.
\newblock Magnetization reversal driven by low dimensional chaos in a nanoscale
  ferromagnet.
\newblock {\em Nature communications}, 10(1):543, 2019.

\bibitem{jenkins2019nanoscale}
Alex~S Jenkins, Lara San~Emeterio Alvarez, Paulo~P Freitas, and Ricardo
  Ferreira.
\newblock Nanoscale true random bit generator based on magnetic state
  transitions in magnetic tunnel junctions.
\newblock {\em Scientific Reports}, 9(1):15661, 2019.

\bibitem{tulapurkar2005spin}
AA~Tulapurkar, Y~Suzuki, A~Fukushima, H~Kubota, H~Maehara, K~Tsunekawa,
  DD~Djayaprawira, N~Watanabe, and S~Yuasa.
\newblock Spin-torque diode effect in magnetic tunnel junctions.
\newblock {\em Nature}, 438(7066):339--342, 2005.

\bibitem{skirdkov2020spin}
Petr~N Skirdkov and Konstatin~A Zvezdin.
\newblock Spin-torque diodes: From fundamental research to applications.
\newblock {\em Annalen der Physik}, 532(6):1900460, 2020.

\bibitem{finocchio2021perspectives}
Giovanni Finocchio, Riccardo Tomasello, Bin Fang, Anna Giordano, Vito
  Puliafito, Mario Carpentieri, and Zhongming Zeng.
\newblock Perspectives on spintronic diodes.
\newblock {\em Applied Physics Letters}, 118(16), 2021.

\bibitem{doi:10.1063/1.3057974}
S.~Yakata, H.~Kubota, Y.~Suzuki, K.~Yakushiji, A.~Fukushima, S.~Yuasa, and
  K.~Ando.
\newblock Influence of perpendicular magnetic anisotropy on spin-transfer
  switching current in cofeb/mgo/cofeb magnetic tunnel junctions.
\newblock {\em Journal of Applied Physics}, 105(7):07D131, 2009.

\bibitem{Ikeda2010}
S.~Ikeda, K.~Miura, H.~Yamamoto, K.~Mizunuma, H.~D. Gan, M.~Endo, S.~Kanai,
  J.~Hayakawa, F.~Matsukura, and H.~Ohno.
\newblock A perpendicular-anisotropy cofeb–mgo magnetic tunnel junction.
\newblock {\em Nature Materials}, 9(9):721--724, 2010.

\bibitem{RevModPhys.89.025008}
B.~Dieny and M.~Chshiev.
\newblock Perpendicular magnetic anisotropy at transition metal/oxide
  interfaces and applications.
\newblock {\em Rev. Mod. Phys.}, 89:025008, Jun 2017.

\bibitem{garello2013symmetry}
Kevin Garello, Ioan~Mihai Miron, Can~Onur Avci, Frank Freimuth, Yuriy
  Mokrousov, Stefan Bl{\"u}gel, St{\'e}phane Auffret, Olivier Boulle, Gilles
  Gaudin, and Pietro Gambardella.
\newblock Symmetry and magnitude of spin--orbit torques in ferromagnetic
  heterostructures.
\newblock {\em Nature nanotechnology}, 8(8):587--593, 2013.

\bibitem{anderson1964lectures}
PW~Anderson.
\newblock {\em Lectures on the Many-body Problem}.
\newblock Academic Press, 1964.

\bibitem{fradkin_2013}
Eduardo Fradkin.
\newblock {\em Field Theories of Condensed Matter Physics}.
\newblock Cambridge University Press, 2 edition, 2013.

\bibitem{PhysRevLett.107.066604}
B.~Heinrich, C.~Burrowes, E.~Montoya, B.~Kardasz, E.~Girt, Young-Yeal Song,
  Yiyan Sun, and Mingzhong Wu.
\newblock Spin pumping at the magnetic insulator (yig)/normal metal (au)
  interfaces.
\newblock {\em Phys. Rev. Lett.}, 107:066604, Aug 2011.

\bibitem{10.1038/s41467-018-05732-1}
Lucile Soumah, Nathanand Beaulieu, Liliaand Qassym, Cécileand Carrétéro,
  Ericand Jacquet, Jamal Lebourgeois, Richardand Ben~Youssef, Paolo Bortolotti,
  Vincent Cros, and Abdelmadjid Anane.
\newblock Ultra-low damping insulating magnetic thin films get perpendicular.
\newblock {\em Nature Communications}, 9:3355, Aug 2018.

\bibitem{BLACKBURN20102827}
James~A. Blackburn, Matteo Cirillo, and Niels Grønbech-Jensen.
\newblock On the classical model for microwave induced escape from a josephson
  washboard potential.
\newblock {\em Physics Letters A}, 374(28):2827--2830, 2010.

\bibitem{BLACKBURN20161}
James~A. Blackburn, Matteo Cirillo, and Niels Grønbech-Jensen.
\newblock A survey of classical and quantum interpretations of experiments on
  josephson junctions at very low temperatures.
\newblock {\em Physics Reports}, 611:1--33, 2016.
\newblock A survey of classical and quantum interpretations of experiments on
  Josephson junctions at very low temperatures.

\bibitem{PhysRevLett.97.050502}
Matthias Steffen, M.~Ansmann, R.~McDermott, N.~Katz, Radoslaw~C. Bialczak, Erik
  Lucero, Matthew Neeley, E.~M. Weig, A.~N. Cleland, and John~M. Martinis.
\newblock State tomography of capacitively shunted phase qubits with high
  fidelity.
\newblock {\em Phys. Rev. Lett.}, 97:050502, Aug 2006.

\bibitem{PhysRevLett.93.107002}
N.~Gr\o{}nbech-Jensen, M.~G. Castellano, F.~Chiarello, M.~Cirillo, C.~Cosmelli,
  L.~V. Filippenko, R.~Russo, and G.~Torrioli.
\newblock Microwave-induced thermal escape in josephson junctions.
\newblock {\em Phys. Rev. Lett.}, 93:107002, Aug 2004.

\bibitem{Landau1978}
L.D. Landau and E.M. Lifshitz.
\newblock {\em Theoretical Physics. Vol. IX. Condensed Matter Theory}.
\newblock M.: Science, 1978.

\bibitem{Beleggia_2005}
M~Beleggia, M~De Graef, Y~T Millev, D~A Goode, and G~Rowlands.
\newblock Demagnetization factors for elliptic cylinders.
\newblock {\em Journal of Physics D: Applied Physics}, 38(18):3333, sep 2005.

\bibitem{zhao2016experimental}
Yuelei Zhao, Qi~Song, See-Hun Yang, Tang Su, Wei Yuan, Stuart~SP Parkin, Jing
  Shi, and Wei Han.
\newblock Experimental investigation of temperature-dependent gilbert damping
  in permalloy thin films.
\newblock {\em Scientific reports}, 6(1):22890, 2016.

\bibitem{10.1063/1.4958855}
N.~Awari, S.~Kovalev, C.~Fowley, K.~Rode, R.~A. Gallardo, Y.-C. Lau, D.~Betto,
  N.~Thiyagarajah, B.~Green, O.~Yildirim, J.~Lindner, J.~Fassbender, J.~M.~D.
  Coey, A.~M. Deac, and M.~Gensch.
\newblock {Narrow-band tunable terahertz emission from ferrimagnetic Mn3-xGa
  thin films}.
\newblock {\em Applied Physics Letters}, 109(3):032403, 07 2016.

\bibitem{PhysRevLett.117.087203}
Takayuki Shiino, Se-Hyeok Oh, Paul~M. Haney, Seo-Won Lee, Gyungchoon Go,
  Byong-Guk Park, and Kyung-Jin Lee.
\newblock Antiferromagnetic domain wall motion driven by spin-orbit torques.
\newblock {\em Phys. Rev. Lett.}, 117:087203, Aug 2016.

\bibitem{PhysRevB.95.144402}
So~Takei, Yaroslav Tserkovnyak, and Masoud Mohseni.
\newblock Spin superfluid josephson quantum devices.
\newblock {\em Phys. Rev. B}, 95:144402, Apr 2017.

\bibitem{PhysRevB.67.094510}
John~M. Martinis, S.~Nam, J.~Aumentado, K.~M. Lang, and C.~Urbina.
\newblock Decoherence of a superconducting qubit due to bias noise.
\newblock {\em Phys. Rev. B}, 67:094510, Mar 2003.

\bibitem{landau2013statistical}
Lev~Davidovich Landau and Evgenii~Mikhailovich Lifshitz.
\newblock {\em Statistical Physics: Volume 5}, volume~5.
\newblock Elsevier, 2013.

\bibitem{PhysRevLett.101.066601}
A.~L. Chudnovskiy, J.~Swiebodzinski, and A.~Kamenev.
\newblock Spin-torque shot noise in magnetic tunnel junctions.
\newblock {\em Phys. Rev. Lett.}, 101:066601, Aug 2008.

\bibitem{PhysRevB.59.2070}
Guido Burkard, Daniel Loss, and David~P. DiVincenzo.
\newblock Coupled quantum dots as quantum gates.
\newblock {\em Phys. Rev. B}, 59:2070--2078, Jan 1999.

\bibitem{coleman_1985}
Sidney Coleman.
\newblock {\em Aspects of Symmetry: Selected Erice Lectures}.
\newblock Cambridge University Press, 1985.

\bibitem{Kockum2019}
Anton~Frisk Kockum and Franco Nori.
\newblock {\em Quantum Bits with Josephson Junctions}, pages 703--741.
\newblock Springer International Publishing, Cham, 2019.

\end{thebibliography}
\end{document}